\documentclass[]{aa}  

\usepackage{graphicx}
\usepackage{txfonts}
\usepackage{CJKutf8} %package for Chinese/Japanese characters
\usepackage{textcomp} % for \AA (Angstrom symbol)
\usepackage{graphicx}
\usepackage{subcaption}
\usepackage{amsmath}

\usepackage[dvipsnames]{xcolor}
\usepackage[colorlinks=true]{hyperref}

\hypersetup{
    colorlinks=true,
    linkcolor=blue,
    filecolor=blue,      
    urlcolor=blue,
    citecolor=teal
}

\newcommand{\Msun}{\ensuremath{\mathrm{M}_\odot}\xspace}

\newcommand{\sfrratio}{\ensuremath{\rm log(SFR_{H\alpha}/SFR_{100Myr})}\xspace}

\newcommand{\forcepho}{\textsc{ForcePho}\xspace}
\newcommand{\ppxf}{ppxf\xspace}
\newcommand{\pccs}{Partial Correlation Coefficients\xspace}

\newcommand{\oiiL}{[O\,\textsc{ii}]\,$\lambda\lambda$}
\newcommand{\oiiiL}{[O\,\textsc{iii}]\,$\lambda$}

\newcommand{\neiiiL}{[Ne\,\textsc{iii}]\,$\lambda$}

\newcommand{\NIV}{N\,\textsc{iv}]\,$\lambda\lambda$1483,1486}

\newcommand{\CIV}{C\,\textsc{iv}\,$\lambda\lambda$1548,1551}

\newcommand{\NeOtwo}{\rm Ne3O2}
\newcommand{\Othreetwo}{\rm O32}

\newcommand{\oasis}[1]{OASIS}

\begin{document}

   % \title{OASIS: Central Starbursts and Inside-Out Quenching?
   % %, and Size Growth 
   % Connecting Bursty Star Formation to Galaxy Sizes at $z>4$}
   \title{OASIS III: Bursty star formation drives observed galaxy sizes at $z\geq4$}
   \titlerunning{High-z burstiness and sizes}

    \author{William M. Baker\inst{1}\thanks{\email{william.baker@nbi.ku.dk}; these authors contributed equally to this work}
\and Tobias J. Looser\inst{2}\thanks{\email{tobias.looser@cfa.harvard.edu}; these authors contributed equally to this work}
\and Francesco D'Eugenio\inst{3,4}
\and Andrew J. Bunker\inst{5}
\and Stefano Carniani\inst{6}
\and St\'ephane Charlot\inst{7}
\and Jacopo Chevallard\inst{5}
\and Mirko Curti\inst{8}
\and A. Lola Danhaive\inst{3,4}
\and Qiao Duan\inst{3,4}
\and Kevin Hainline\inst{9}
\and Jakob M. Helton\inst{10}
\and Benjamin D. Johnson\inst{2}
\and Pierluigi Rinaldi\inst{11,12}
\and Brant Robertson\inst{13}
\and Fengwu Sun\inst{2}
\and Yang Sun\inst{9}
\and Sandro Tacchella\inst{3,4}
\and James A. A. Trussler\inst{2}
\and Zihao Wu\inst{2}
}
 
\institute{DARK, Niels Bohr Institute, University of Copenhagen, Jagtvej 155A, DK-2200 Copenhagen, Denmark
\and Center for Astrophysics $|$ Harvard \& Smithsonian, 60 Garden St., Cambridge, MA 02138, USA
\and Kavli Institute for Cosmology, University of Cambridge, Madingley Road, Cambridge, CB3 0HA, UK
\and Cavendish Laboratory, University of Cambridge, 19 JJ Thomson Avenue, Cambridge, CB3 0HE, UK
\and Department of Physics, University of Oxford, Denys Wilkinson Building, Keble Road, Oxford OX1 3RH, UK
\and Scuola Normale Superiore, Piazza dei Cavalieri 7, I-56126 Pisa, Italy
\and Sorbonne Universit\'e, CNRS, UMR 7095, Institut d'Astrophysique de Paris, 98 bis bd Arago, 75014 Paris, France
\and INAF -- Osservatorio di Astrofisica e Scienza dello Spazio, Via P. Gobetti 93/3, I-40129 Bologna, Italy
\and Steward Observatory, University of Arizona, 933 N. Cherry Avenue, Tucson, AZ 85721, USA
\and Department of Astronomy \& Astrophysics, The Pennsylvania State University, University Park, PA 16802, USA
\and Department of Astronomy, The University of Texas at Austin, Austin, TX 78712, USA
\and Cosmic Frontier Center, The University of Texas at Austin, Austin, TX 78712, USA
\and Department of Astronomy and Astrophysics, University of California, Santa Cruz, 1156 High Street, Santa Cruz, CA 95064, USA
}
   %\subtitle{I. Overviewing the $\kappa$-mechanism}
%    \author{William M. Baker
%           \inst{1}\fnmsep\thanks{william.baker@nbi.ku.dk}
%           \and
%           Tobias J. Looser \inst{2}\fnmsep\thanks{tobias.looser@cfa.harvard.edu
% }
%           \and
%           Francesco D'Eugenio
%           \and
%           et al.
%           }

%    \institute{DARK, Niels Bohr Institute, University of Copenhagen, Jagtvej 155A, DK-2200 Copenhagen, Denmark 
%         \and
%         Center for Astrophysics $\|$ Harvard \& Smithsonian, 60 Garden St., Cambridge, MA 02138, USA
%         }

   % \date{Received September 15, 1996; accepted March 16, 1997}

% \abstract{}{}{}{}{} 
% 5 {} token are mandatory
 
  \abstract
  % context heading (optional)
  % {} leave it empty if necessary  
   {
   We employ \emph{JWST}/NIRSpec MSA PRISM spectroscopy together with NIRCam imaging to investigate the impact of star-formation variability on galaxy sizes for a representative sample of 51 galaxies at $z\geq4$. The OASIS survey enables us to probe low-star-formation-rate (SFR) \emph{lulling} and \emph{mini-quenched} galaxies down to $\log_{10}(M_\star/M_\odot)=8.3$ and $\log_{10}(\mathrm{SFR}/M_\odot\,\mathrm{yr}^{-1})<-1.6$, i.e. more than one order of magnitude below the star-forming main sequence (MS). This enables us to probe objects that are often missed by shallower spectroscopic surveys, in addition to the MS and starburst systems, yielding an SFR- and mass-representative sample at high redshift. We derive galaxy half-light sizes using \textsc{ForcePho}, enabling us to resolve the highly compact nature of these sources, and we analyse their star-formation histories and interstellar medium (ISM) properties with the full-spectral-fitting code \ppxf. We find a clear anti-correlation between burstiness, as traced by the ratio of the SFR from H$\alpha$ compared to that from non-parametric spectral fitting averaged over 100Myr, and galaxy UV/optical half-light radii as measured with \forcepho from a combination of well-sampled NIRCam photometry. 
   Using partial correlation coefficients we show that the anti-correlation between burstiness (as traced by the SFR ratio) and size cannot simply be explained by secondary dependencies on stellar mass or redshift. We interpret this relation as the  effect of \emph{central starbursts} in SFR-rising (bursting) systems, where intense central star formation reduces measured galaxy half-light sizes. This burstiness then regulates galaxy sizes at high redshift. Should the same mechanism be present at $z>10$, this could explain the size bimodality seen in the UV luminous galaxies as simply stemming from a more fundamental relationship between burstiness and size.
   This result has important implications for our understanding of burstiness in high-z galaxy evolution and also strongly motivates incorporating galaxy sizes into completeness calculations when exploring population-wide analyses.

   }

   \keywords{Galaxies: high-redshift, Galaxies: evolution, Galaxies: formation, Galaxies: star-formation}

   \maketitle
%
%-------------------------------------------------------------------

\section{Introduction}

Galaxies at high redshift ($z>4$) and low masses ($\log(M_\star/{\rm M_\odot})\lesssim 9$) are subject to strong variability (burstiness) in their star-formation histories (SFHs). This is likely caused by the complex interplay between gas accretion, star formation, and feedback \citep[e.g.][]{Hopkins2014, El-Badry2016,Tacchella2016, McClymont2025b}. Galaxies accrete fresh gas that fuels star formation, rapidly converting it into stars, until stellar or AGN feedback suppresses further activity \citep[e.g.][]{Dekel1986,Fabian2012,Croton2006, Hayward2017}. Feedback can expel the gas or prevent it from cooling efficiently, temporarily suppressing subsequent star formation. 

These stochastic processes, and their interplay, cause oscillations around the `Star-forming Main Sequence' (hereafter MS), a relation between a galaxy's stellar mass and its associated star-formation rate (SFR) \citep[][]{Brinchmann2004, Noeske2007,Speagle2014, Tacchella2016, Simmonds2025, Rinaldi2025,Clarke2025, Cole2025}, itself a by-product of the connection between SFR, stellar mass, and gas mass \citep[at least up to $z\sim3$,][]{Baker2023}. Hence, these stochastic modulations are associated with the duty cycle of these high-redshift galaxies, their feedback, and gas-accretion processes \citep{Tacchella2016, Gelli2025}. 
Stochasticity is predicted by cosmological simulations and semi-analytic models to become more pronounced at higher redshifts and lower masses, due to effects such as shallower gravitational potentials, shorter dynamical timescales, and highly fluctuating gas accretion rates \citep{Tacchella2016b,Dome2024, Gelli2023,Sun2023b, Gelli2025, Kravtsov2024, Munoz2026}. This prediction of increased stochasticity appears consistent with high-z observations \citep[e.g.][]{Endsley2024,Endsley2025,Boyett2024,Langeroodi2024,Looser2025, Perry2025, Clarke2025, Cole2025, Simmonds2025,Navarro-Carrera2026}. 

Indeed, within its first years of operation, JWST has discovered numerous UV-bright galaxies at $z>4$ (and up to $z\sim14$) that appear to be experiencing intense starburst episodes at the time of observation \citep[e.g.][]{Curtis-Lake2023b,Bunker2023b,Carniani2024, Castellano2024, Naidu2025, Donnan2026}. However, this view is strongly affected by selection biases, especially in the high-redshift and low-mass regime: spectroscopic programmes preferentially target the brightest sources, while fainter galaxies, particularly in shallow surveys, remain under-explored. Comparisons between spectroscopic samples and both mass-complete photometric studies \citep[e.g.][]{Simmonds2024,Endsley2024,Endsley2025}, and simulations \citep[e.g.][]{Sun2023b, Gelli2025} highlight this discrepancy, indicating that low-mass, low-SFR, intrinsically faint systems, likely in an “off” phase of their duty cycle, are systematically missing from spectroscopic datasets.

Burstiness has hence been proposed as a potential solution to the well-known overabundance of UV-bright galaxies at high redshift \citep[e.g.][]{Mason2023, Sun2023a, Mitsuhashi2026}. In this picture, observations are biased toward galaxies caught in luminous burst phases, while fainter, lower-SFR counterparts are under-represented. Nevertheless, other recent work suggests that potentially burstiness alone cannot provide a complete explanation for this UV bright galaxy overabundance \citep{Gelli2024b,Simmonds2025, Kokorev2025_capers, WeibelJesp2025, Kreilgaard2026}.

Despite being biased towards burst-phase galaxies, we are now assembling spectroscopic confirmation of a handful of so called `mini-quenched' galaxies \citep{Looser2024, Baker2025b} and low-SFR `lulling' galaxies \citep{Strait2023,Looser2025, Covelo-Paz2025,Harikane2026, Khullar2026}. These faint, low-mass galaxies show clear Balmer breaks, relatively steep UV slopes, and a lack of strong emission lines, all of which is consistent with a picture in which these galaxies have very recently (either completely or nearly completely) ceased forming new stars, likely within the previous 10-100~Myr. This neatly explains the observed properties of these galaxies: no strong line emission, due to a lack of star formation within the previous $\approx$10~Myr \citep{Looser2024, Baker2025b}, a visible Balmer break, due to an absence of the hottest and youngest stars \citep{Sawicki1998, Looser2025,Witten2025}, and an often blue UV slope due to a young stellar population of 50-150 Myr \citep{Wilkins2011} \citep[][hereafter OASIS Paper~I]{Looser2026a}. \footnote{It can be most convenient to think of the `mini-quenched' galaxies as a separate population compared to the more traditional definition of a `massive quiescent' galaxy. The `massive quiescent' galaxies are typically selected via either UVJ colour selection \citep[e.g][]{Schreiber2018} or a specific star-formation rate (sSFR) cut averaged over 100Myr \citep[e.g.][]{Franx2008, Carnall2018, Baker2025d}. Mini-quenched galaxies do not usually fulfil either of these criteria, as they are much bluer and lower mass ($<10^9 \Msun$).
For a detailed comparison of low-mass quiescent galaxies vs mini-quenched galaxies see \citet{Baker2026}.}

However, in order to constrain the star-formation burstiness further, we need to probe more than the single galaxy samples in \citet{Looser2024} and \citet{Baker2025b}. While photometry can provide large and complete samples \citep[e.g.][]{Trussler2025}, in order to accurately confirm the nature of these galaxies, remove interlopers and constrain their SFHs, we require a spectroscopic sample that specifically targets these `mini-quenched' or `lulling' galaxies. 

The \oasis{} survey (\textit{Observing All phases of StochastIc Star formation}; GO~5997; PI: T.~J.~Looser, co-PI: F.~D’Eugenio; OASIS Paper~I) was specifically designed to provide such a sample; this is the third of three companion papers, with OASIS Paper~I and \citet[][hereafter OASIS Paper~II]{Looser2026b}. Its selection was optimised to probe the full range of MS galaxies, providing a representative sample of galaxies above and below the MS at $z>4$. This enables a more comprehensive spectroscopic analysis on a sample that is not just dominated by galaxies undergoing burst phases. 

A detailed analysis of star-formation burstiness, based on the OASIS sample, including its mini-quenched and lulling galaxies, the scatter around the main sequence, stellar populations, and star-formation histories, has been presented in OASIS Papers~I and II. In this paper, we instead focus on the relationship between galaxy burstiness and size. 

Observationally we measure galaxy sizes based on the half-light radii of galaxies, which gives a measure of how the light is distributed within a galaxy, such as whether it is centrally concentrated in a core or bulge \citep[e.g.][]{vanDokkum2014,Tacchella2015, Nelson2014, Baker2025}, or in an extended disc-like structure \citep[e.g.][]{Smit2018,Danhaive2025, Nelson2024, Baker2026b}.
Predictions from simulations show that, as galaxies oscillate about the main sequence, this should be reflected in their measured sizes \citep{Zolotov2015, Tacchella2016b, McClymont2025}. This is primarily due to two effects. The first of these is a central starburst, likely caused by a compaction of gas within the central region \citep{Dekel2014, Zolotov2015}. The galaxy builds up a central gas rich starburst that rapidly forms stars \citep[whilst the gas accretion rate can counter the associated stellar feedback][]{Tacchella2016b}, causing a centrally concentrated light profile, shrinking the observed half-light radius. This compaction driven central starburst effect is seen clearly in simulations \citep[e.g.][]{Zolotov2015,McClymont2025}.
Also note that this effect is separate from the more steady inside-out growth mechanisms, seen at cosmic noon \citep[e.g.][]{vanDokkum2010, Nelson2016}, whose prevalence is under debate at higher redshift \citep{Matharu2024, Danhaive2026, Baker2025}.
% This is in agreement with many observations of inside-out growth around cosmic noon \citep{vanDokkum2010,vanDokkum2014,Nelson2016} and findings of inside-out growth and central starbursts in the early Universe \citep{Baker2025, Matharu2024}. 

The second potential effect relates to the quenching, i.e. the shutdown, of star-formation within these galaxies \citep[for an observational review see][]{Bluck2026}. Depending on the quenching process, different parts of a galaxy may quench first, in a process known as inside-out or outside-in quenching which could affect the measured size of galaxies \citep[e.g.][]{Zolotov2015, Tacchella2015}.

In this paper, we combine OASIS spectroscopy with JADES \citep{Eisenstein2023} imaging to investigate whether burstiness is linked to galaxy size at $z>4$. We utilise the SFR ratio (\sfrratio) as a burstiness indicator, and we assess whether any apparent relation with size remains after controlling for both stellar mass and redshift. Our target is to determine whether compact starbursts and more extended lulling/mini-quenched phases trace a common structural cycle in high-z galaxy evolution.
Throughout this paper we use a \citet{Chabrier2003} Initial Mass Function and \citet{PlanckCollaboration2020} cosmology.

\section{Data and Methods}
\label{sec:data}

This work combines NIRSpec MSA prism spectroscopy from the \oasis{} programme
with NIRCam multi-band photometry from the JADES programme \citep[JWST Advanced Deep Extragalactic Survey;][]{Eisenstein2023, Rieke2023}.
The OASIS observations consist of two 28-hour pointings, centred on the JADES Deep pointing in the Hubble Ultra Deep Field \citep[HUDF][]{Beckwith2006} and on the JADES Origins Field \citep{Eisenstein2023jof}.
For the data reduction, we use the JADES spectroscopic and photometric reduction pipelines \citetext{respectively \citealp{Scholtz2025} and Tacchella et~al., in prep}. The full sample of OASIS galaxies is presented in OASIS Paper~I. 
%Two example OASIS spectra corresponding to a lulling galaxy (left) and a starburst galaxy (right) are presented in Fig. \ref{fig:example_spectra}.

To measure galaxy sizes, we forward model the observed light distribution in each individual exposure using the code \forcepho \citep[Johnson et al., in prep; see][for more details]{Robertson2023, Baker2025a}. \forcepho fits the observed light distribution in each individual exposure, thus enabling accurate sub-pixel resolution by fully leveraging the dithering. Fluxes are allowed to vary between the filters, whilst the half-light radius and other structural parameters are fit based on the information in all exposures. We follow the setup of \citet{Baker2025} and model all galaxies and neighbouring sources with single S\'ersic profiles. This means that throughout this work, size will refer to the continuum half-light radius from single S\'ersic \forcepho fits performed jointly across all available imaging and therefore traces the spatial concentration of observed light rather than a mass radius. Whilst \forcepho enables us to take into account dithering to obtain a sub-pixel resolution, in order to take into account point sources, we clip sizes smaller than 0.015\arcsec (half a SW pixel size in the JADES data reduction) to this value. This does not change any of our findings.

The spectra are fit with the spectral fitting code \textsc{ppxf} \citep{Cappellari2017, Cappellari2023}. The full spectral fitting routine is presented in detail in OASIS Paper~I. Our burstiness indicator is the ratio of the SFR from dust-corrected H$\alpha$ (using the \citet{Shapley2023} calibration for metal-poor stellar populations; OASIS Paper~I), which roughly traces 10Myr timescales, to the SFR from the full spectral fitting averaged over 100Myr. The full derivations, details and tests regarding these properties can be found in OASIS Papers~I and II. We will also split the galaxies into a bursting and declining group based on whether \sfrratio is greater or lower than 0. This also roughly corresponds to rising or falling SFHs respectively.

\section{Results}

\subsection{Overall Sample}

\label{s.sample}

\begin{figure}
    \centering
    \includegraphics[width=1\linewidth]{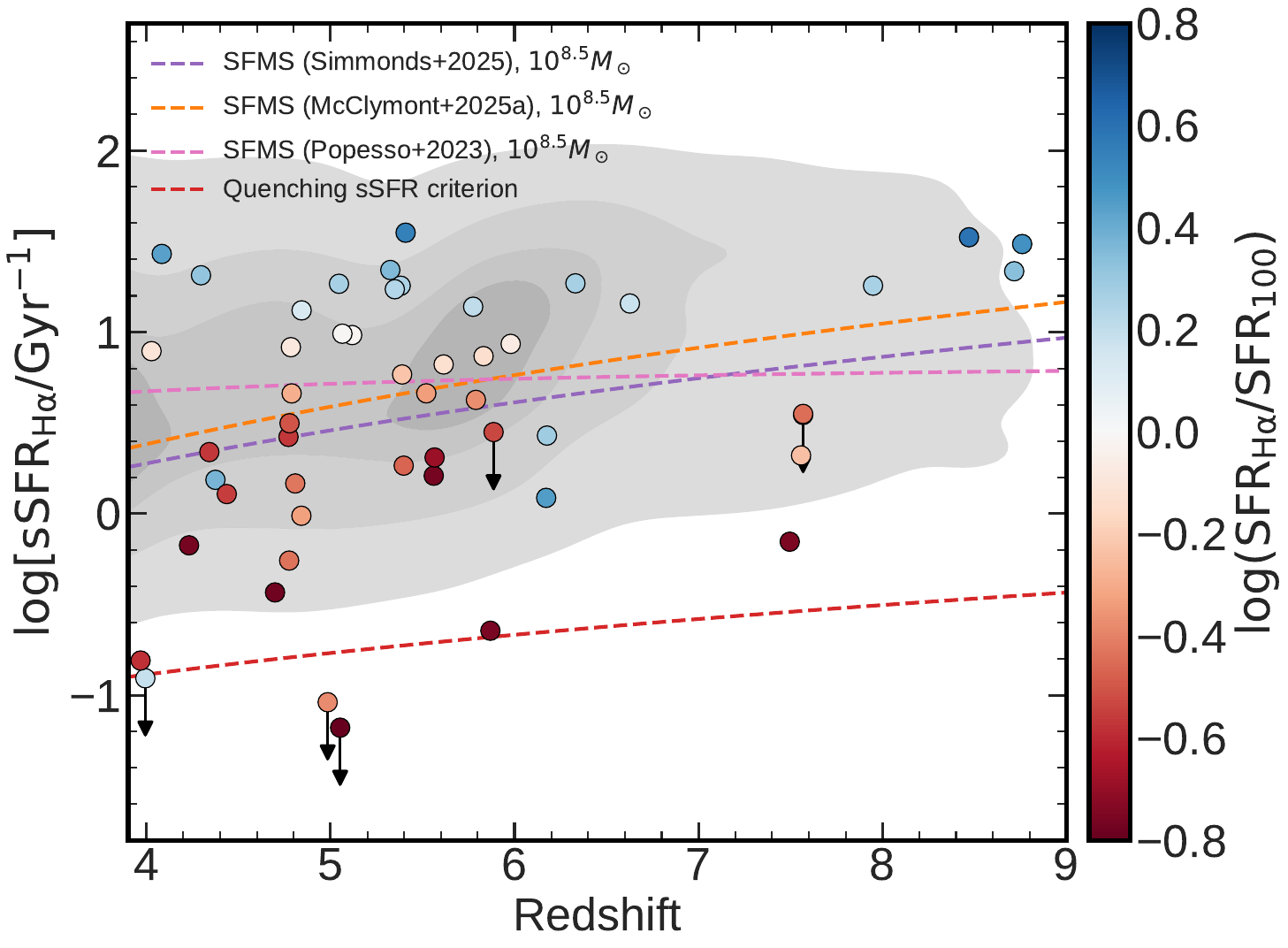}
    \caption{Specific star-formation rate vs redshift for our sample (coloured points, $3\sigma$ upper limits are denoted via downward arrows) colour-coded by \sfrratio. MS from the literature are overplotted \citep{Popesso2023,Simmonds2025, McClymont2025} alongside a typical quenching threshold. The mass-complete photometric sample of \citet{Simmonds2025} is plotted as grey contours. It shows that the OASIS sample probes galaxies on, above and below the MS and is a representative sample for our analysis.}
    \label{fig:ssfr_redshift}
\end{figure}

Our sample is selected to probe the $z>4$ galaxy population on, above and below the MS. Hence, we apply a redshift cut of $z>3.95$ (following OASIS Paper~I) to the overall OASIS sample based on the spectroscopic redshifts. This leaves us with 64 galaxies.
We remove 5 galaxies poorly modelled by ppxf, including  
4 with low continuum S/N leaving the stellar mass and SFH unconstrained (216135, 119154, 307492, and 308149) and 1 (165902) due to being a broad line AGN with S/N$\;\sim190$. 
  
In addition, we require that the galaxies' light distribution is well reproduced by \forcepho. Hence, we visually inspect every resulting \forcepho fit and remove obvious poor fits caused by mergers or particularly clumpy galaxies (see Sec. \ref{a.s.forcepho} for more details). This results in the exclusion of 5 galaxies whose fits are shown in Appendix \ref{a.s.forcepho}.
In addition, OASIS targeted three $z>10$ galaxies which we remove from the sample due to the difficulty of robustly constraining their stellar population properties and star-formation histories (particularly their \sfrratio ratio due to lacking H$\alpha$ or H$\beta$ coverage).

Next, we explore our final sample of galaxies. This consists of 51 galaxies within the redshift range $z\approx4-9$, probing some of the earliest galaxies from just $\sim$500Myr after the start of the Universe.

Fig. \ref{fig:ssfr_redshift} shows the distribution of these galaxies in relation to the MS with SFRs calculated from emission lines following OASIS Paper~I. The coloured points correspond to the sample, with downwards arrows representing SFR $3\sigma$ upper limits. Overplotted are the MS prescriptions of the pre-JWST \citet{Popesso2023}, the simulation based \citet{McClymont2025}, and the JWST photometric work of \citet{Simmonds2025}. The grey contours correspond to the distribution of galaxies based on the mass-complete photometric sample in \citet{Simmonds2025}.
The red line corresponds to a mini-quenched criterion of
\begin{equation}
   \rm sSFR_{em} \leq \frac{0.2}{t_{age}},
\end{equation}
where $\rm t_{age}$ is the age of the Universe at the observed redshift. This criterion has been used extensively throughout the literature to identify quiescent galaxies \citep[e.g.][]{Franx2008, Pacifici2016,Carnall2018, Baker2025a, Baker2025d}, when based on the SFR averaged over 100~Myr. In this case we adapt this criterion such that it can be used for mini-quenched galaxies.
As our SFRs are from emission lines, it represents a helpful selection of mini-quenched galaxies that have not had considerable star-formation within the last 10-20~Myr. 

The key takeaway from Fig. \ref{fig:ssfr_redshift} is that the \oasis{} sample traces well the population on, above and below the MS, providing an ideal sample to probe all phases of bursty star formation. We see that this is the case regardless of the MS definition used. Whilst the vast majority of our galaxies lie in the redshift range $z=4-7$, even the higher redshift galaxies seem well-distributed around the MS. 
This broad coverage of above, on, and below the MS is what enables a morphological test of bursty star-formation that prior spectroscopic samples could not easily perform.

Of the final sample of 51 galaxies, 46 have a detected Balmer line from which we
measure SFR$_{\mathrm{H}\alpha}$ and hence $\log(\mathrm{SFR_{H\alpha}/SFR_{100Myr}})$.
39 of these via H$\alpha$ directly and 7 at $z>7.1$ (where H$\alpha$ is no longer covered) via H$\beta$ (see OASIS Paper~I). The
remaining 5 have a 3$\sigma$ H$\alpha$ (4) or H$\beta$ (1) upper limit, hence the \sfrratio is also an upper limit. All
correlation analyses (Figs.~\ref{fig:size_burstiness}, \ref{fig:pcc_arrow}, \ref{fig:pcc_bar} and Fig. \ref{fig:plane}, Table~\ref{tab:sizemass}) use the 46
H$\alpha$/H$\beta$-detected galaxies, with the upper limits shown only for reference. For the analysis using stacking (Fig. \ref{fig:light_profiles}) we divide the sample into bursting (\sfrratio$>0$ with 18
galaxies) and declining (\sfrratio$\leq0$ with 31 galaxies, including the 3 H$\alpha$ upper limits
whose \sfrratio is constrained to be $<0$). The 2 upper limits with \sfrratio$>0$ are excluded (as they cannot be added to either group), as is one bursting galaxy from the
F200W stack, which lacks F200W coverage. 
Analyses requiring additional
measurements use the following subsets: [Ne\,{\sc iii}]/[O\,{\sc ii}] and
[O\,{\sc iii}]/[O\,{\sc ii}] are detected in 30 and 38 galaxies respectively (Fig.~\ref{fig:size-U}) and Balmer
decrements are measurable in 37 galaxies (Fig.~\ref{fig:dust-size}).
For the five galaxies with ForcePho sizes that we treat as upper limits (at the $0.015''$), we find that excluding them does not change our conclusions.

\subsection{Effect of selection}

Naturally in any spectroscopic based attempt to understand galaxy populations a key caveat is selection effects and we note these here. With MSA slit spectroscopy we only obtain a sample of certain selected targets. 
Galaxies below the MS that are not yet quiescent \citep[e.g.][]{Looser2024,Strait2023, Baker2025b, Covelo-Paz2025} are a particularly challenging population to study in light of the trade off between completeness and purity.
Although OASIS deliberately targeted a combination of galaxies above, on and below the MS, and hence the sample studied in this work should be representative, it cannot be complete. For more details about the exact completeness and selection of galaxies within the OASIS survey we refer readers to OASIS Paper~I.

The usual ways to ensure a complete sample are to use photometric based samples \citep[e.g.][]{Trussler2025} or alternative approaches such as slitless spectroscopy which can target all galaxies within a given field \citep[e.g.][]{Danhaive2025}. 
However, neither of these two approaches work for the lulling or mini-quenched galaxy populations, and hence either would severely bias our analysis (see OASIS Paper~I). In the case of slitless spectroscopy these galaxies lack the strong emission lines that would be required to detect them in the NIRCam grism making it physically impossible, while photometry is unable to accurately distinguish lulling or mini-quenched galaxies from actively star-forming galaxies. 

The OASIS sample itself particularly highlights this, as explained in OASIS Paper~I. Amongst the ten highest quality mini-quenched or lulling candidates as selected from well-sampled photometry, 7 were found to be star-forming, whilst conversely several mini-quenched or lulling galaxies were found out of the low-priority targets (OASIS Paper~I). Were this a purely photometric study, this would significantly bias our findings due to being unable to accurately determine the lulling or mini-quenched galaxy populations. Our spectroscopic approach therefore mitigates this problem, albeit with the key caveat that our sample cannot be complete.
In order to study a complete population accurately we will require a full complete spectroscopic survey as photometry alone is clearly not enough to accurately select these galaxies.

\subsection{Burstiness vs size}

\begin{figure}
    \centering
    \includegraphics[width=0.99\linewidth]{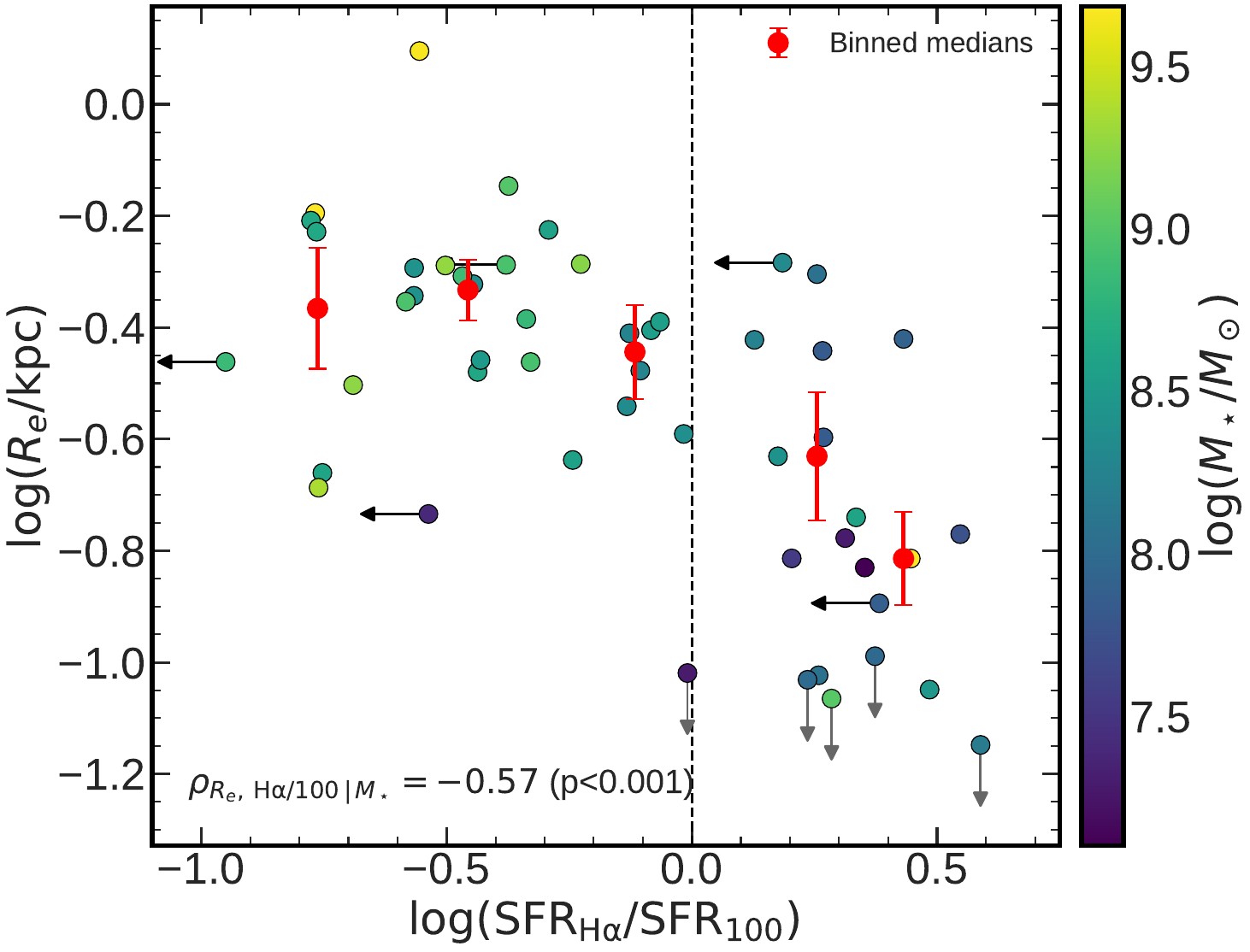}
    \caption{Effective half-light radius from \forcepho multi-band fitting vs burstiness (\sfrratio) colour-coded by stellar mass. The red points correspond to the bin medians, with the errors corresponding to the standard error on the mean. It shows a clear relation between burstiness and the size as measured by the half-light radius. The partial correlation coefficients between size and \sfrratio show a statistically significant anti-correlation of -0.57.
     Galaxies with greater burstiness (higher \sfrratio) are systematically smaller.}
    \label{fig:size_burstiness}
\end{figure}

\begin{figure*}
    \centering
    \includegraphics[width=0.49\linewidth]{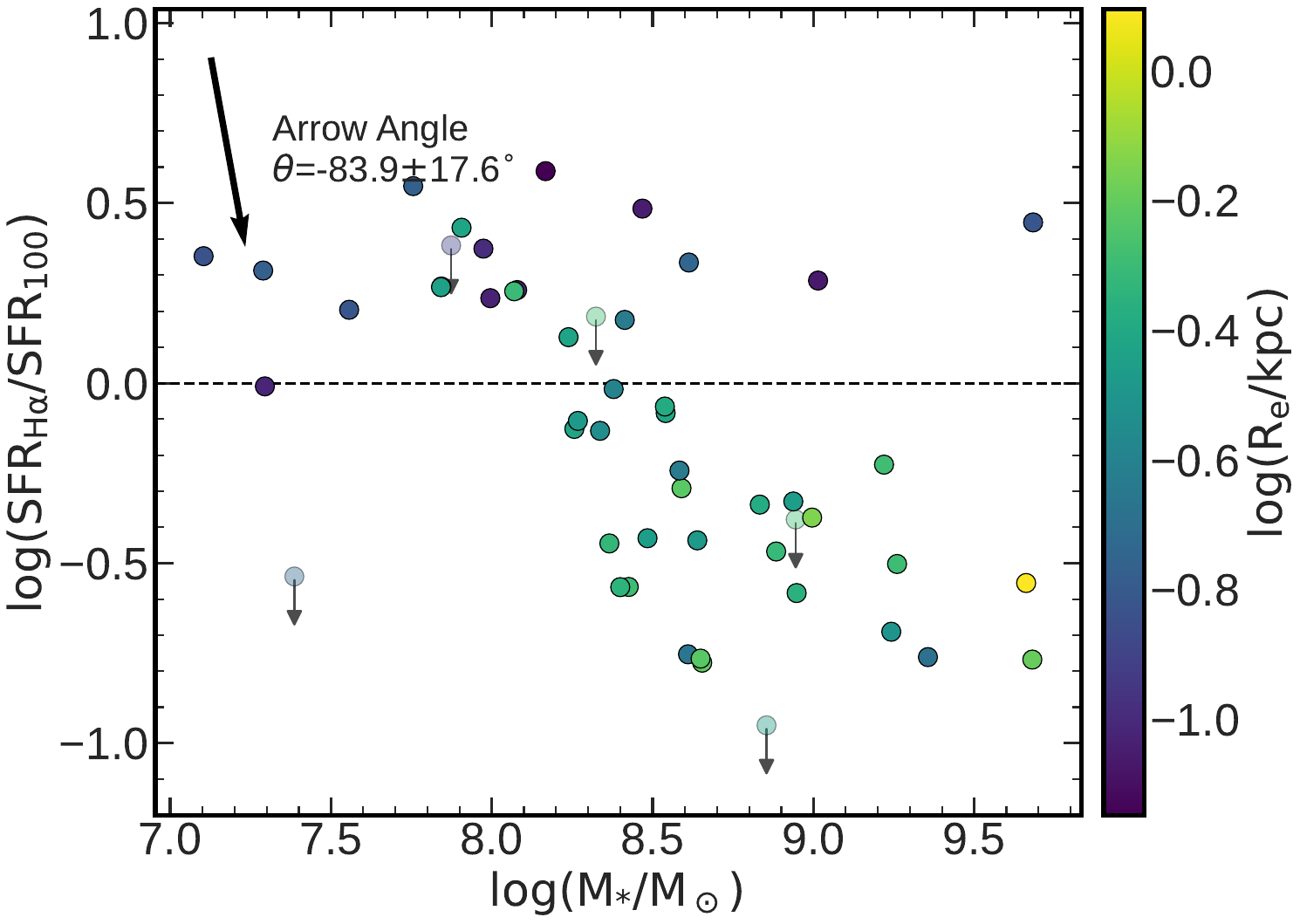}
    \includegraphics[width=0.49\linewidth]{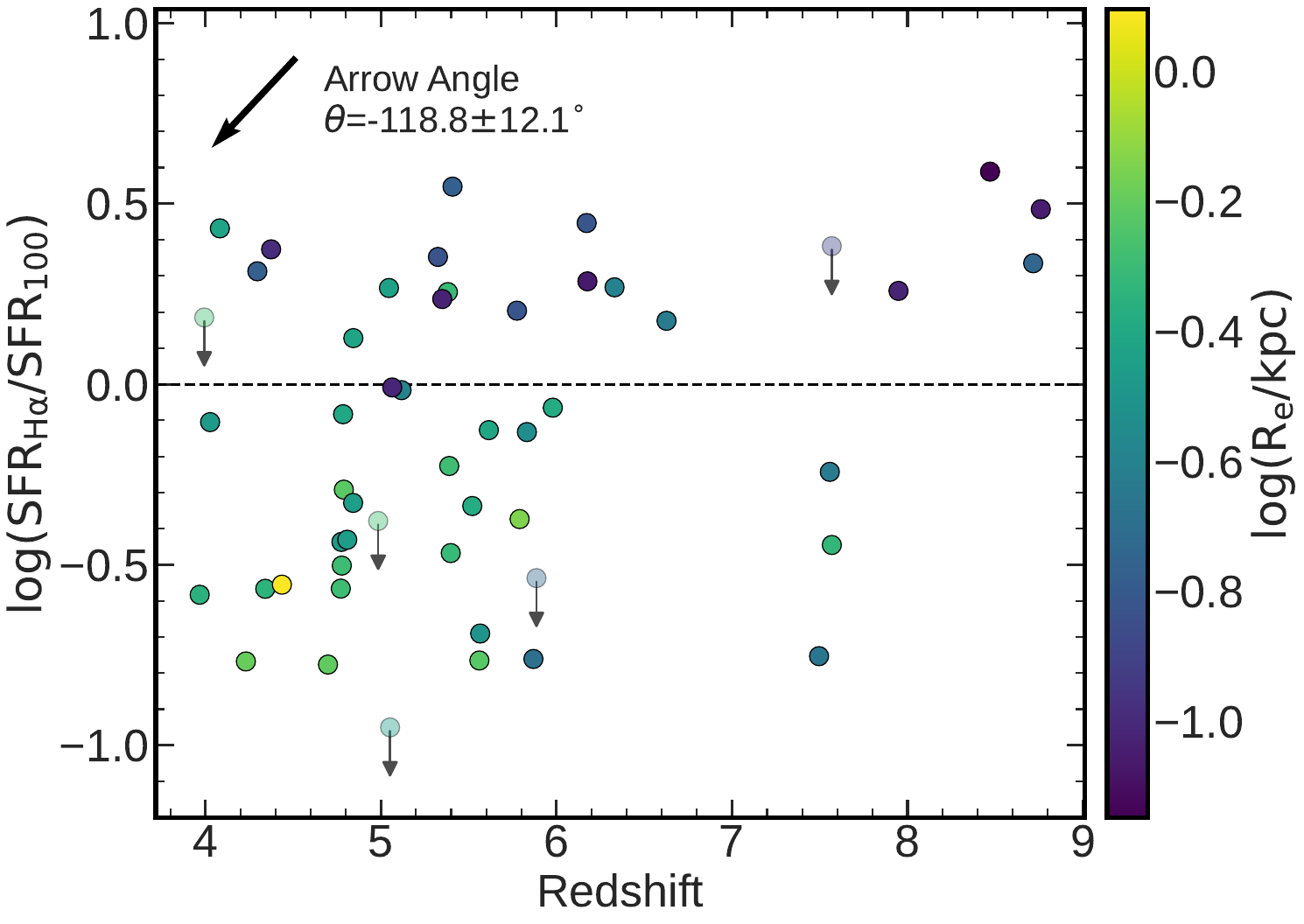}
    \caption{\sfrratio vs stellar mass colour coded by half-light radius (left), and burstiness vs redshift colour-coded by half-light radius (right). The arrow points in the direction of the greatest increase in the colour coded quantity, in this case size \citep[for more details see][]{Baker2022}. The left hand figure shows that size depends primarily on \sfrratio (inversely) and is almost independent of mass, whilst the right hand figure shows that size depends on both burstiness and redshift.}
    \label{fig:pcc_arrow}
\end{figure*}
We now explore the relationship between burstiness and galaxy sizes.
Measuring burstiness is complicated as every measure has its advantages and disadvantages. 

The primary measure we use is the SFR ratio \sfrratio which corresponds to the ratio between the average SFR in the 10-20Myr prior to observation, divided by the average SFR within 100Myr prior to observation. Galaxies with \sfrratio$>0$ are in a burst phase, whilst galaxies with \sfrratio$<0$ are in a declining phase. This provides a simple measure of burstiness. The downside of this approach is that at higher redshifts galaxies can start to show mostly rising SFHs that are not necessarily driven by bursts, rather by continual growth. Another way of putting this is that as opposed to the average SFH being (generally) relatively constant as we approach the highest redshifts, the average SFH should be rising, as expected from the increasing rate of accretion of baryons onto dark-matter haloes \citep[e.g.,][]{Dekel2013,Tacchella2018b}.

Fig. \ref{fig:size_burstiness} provides the most straightforward visualisation of the potential trends between size and burstiness. It shows the half-light effective radius of the galaxies against their \sfrratio colour coded by stellar mass. The red points correspond to binned trends in the data with the errors given by the standard error on the mean. We see a clear trend between the size of a galaxy and its \sfrratio ratio. The greater the \sfrratio ratio, the smaller the galaxy. We also see that this is primarily driven by the bursting galaxies, i.e. those with \sfrratio>0, rather than the declining galaxies which seem to have consistent sizes, regardless of how sharp the decline is.

However, it is natural to consider the effect of stellar mass on these findings. We expect a relationship between mass and burstiness, and size and mass; therefore, the observed trend between size and burstiness could just be a by-product \citep[see][for a clear example of this kind of effect]{Baker2022}. To rule this out, we need to control for the stellar mass of each galaxy as well. The optimum way to do this would be based on a machine learning approach like a random forest regression \citep[e.g.][]{Bluck2022, Baker2022}, but due to our small sample size, a machine learning approach such as this is unfeasible. Therefore, we use an approach based on Partial Correlation Coefficients \citep{Bluck2020,Baker2022}. The partial correlation coefficient between A and B whilst controlling for C is given by \citep{Bluck2020, Baker2022}
\begin{equation}
    \rho_{AB|C}=\frac{\rho_{AB}-\rho_{AC}\rho_{BC}}{\sqrt{1-\rho_{AC}^2}\sqrt{1-\rho_{BC}^2}}.
\end{equation}

\pccs enable the correlation between two quantities to be determined, whilst controlling for a single further quantity \citep{Baker2022}, or for multiple further quantities \citep{Baker2023b}.
We use them for determining the correlation between \sfrratio ratio and size, whilst controlling for the stellar mass. As shown in Fig. \ref{fig:size_burstiness} we find $\rho=-0.57$, with a p-value of $<$0.001, enabling us to robustly reject the Null hypothesis and quantitatively demonstrate the link between burstiness and size. 

The next stage is to investigate the strength of this potential relation. We do this by utilising the ratio of the partial correlation coefficients of the two axis quantities vs the colour coded quantity whilst controlling for the other third quantity \citep{Bluck2020, Baker2022} in Fig. \ref{fig:pcc_arrow}. For more details on this methodology, see \citet{Baker2022}. In short, the arrow points in the direction that best matches the colour shading gradient, i.e. the direction of the arrow tells one how to increase the value of the colour coded quantity, which has the effect of showing its relative dependencies on the x- and y axis quantities. In our case, the colour-coded quantity is the half-light radius.

Fig. \ref{fig:pcc_arrow} shows the results of this methodology. It shows \sfrratio ratio vs stellar mass (left) and redshift (right) colour-coded by size. The arrow angle shows the direction one would travel in to increase the size of the galaxy. 
The left hand panel shows that to increase the size one would need to decrease the \sfrratio (i.e. the colour shading is predominantly vertical),
but that it is essentially independent of mass, i.e. changing the mass at a fixed burstiness would not alter your size.
This enables us to determine which out of mass and burstiness drives the size and we find it is almost entirely burstiness. Despite this, we know that redshift is likely to also contribute to galaxy size \citep[e.g.][]{Mo1998,Ferguson2004}.
The right hand diagram shows that to increase the size, one would both decrease the redshift and lower \sfrratio, almost in equal proportions.  
This shows that the key drivers of size appear to be a combination of \sfrratio and redshift.
In the next section, we explore the results of controlling for both mass and redshift at the same time.

\subsection{Controlling for both stellar mass and redshift}

\begin{figure}
    \centering
    \includegraphics[width=0.90\linewidth]{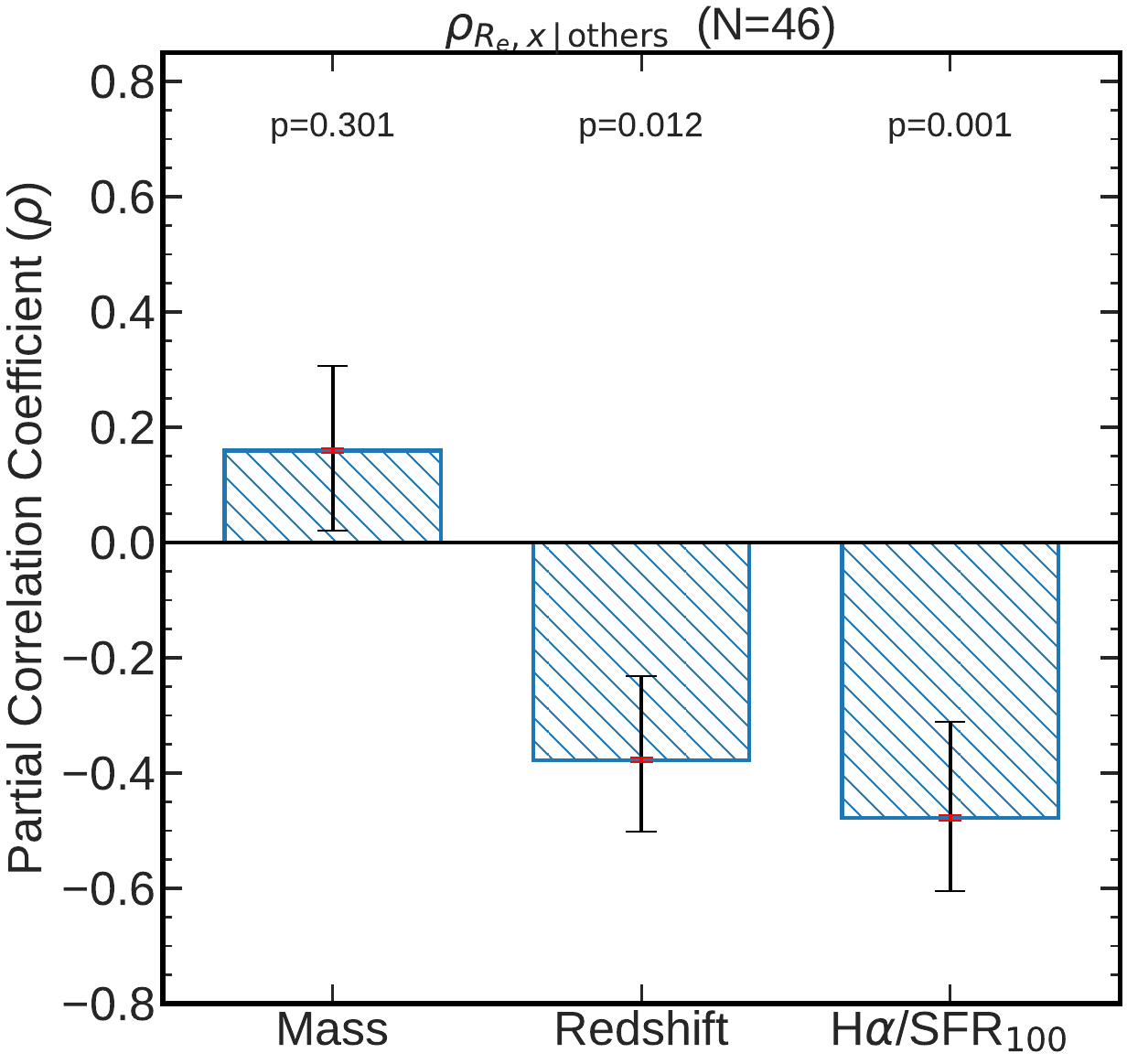}
    \caption{Partial correlation coefficients between all quantities and the size whilst controlling for all other quantities in the figure. The black error bars correspond to bootstrap random sampling 1000 times and taking the 16th and 84th percentiles of the resulting distribution, whilst the red error bars are taking the standard error on the mean. 
    The strongest (anti-)correlation is with the \sfrratio ratio, followed by the redshift, which are the only ones significant enough to reject the null hypothesis. This highlights the strong anti-correlation between \sfrratio and galaxy sizes, even whilst controlling for other variables.}
    \label{fig:pcc_bar}
\end{figure}

In this section, to test whether the size–burstiness relation is intrinsic, rather than a by-product of the size–mass relation or its redshift evolution, we compute partial correlation coefficients, whilst controlling for both stellar mass and redshift.
The difficulty is that the more variables held constant, the larger the required sample size in order to gain significance \citep[][]{Baker2023b}. For that reason we will limit our test to stellar mass, redshift and \sfrratio.

Fig. \ref{fig:pcc_bar} shows the partial correlation coefficients between each of the quantities vs size whilst controlling for the other two quantities in the figure. The y axis shows the strength of the correlation or anti-correlation. The black error bars are obtained from the 16th and 84th percentiles of the bootstrap random sampled distribution (which was bootstrapped 1000 times), whilst the red errors correspond to the standard error on the mean of the bootstrapped distribution. This helps ensure that the errors on the partial correlation coefficients better represent the underlying distribution. 

The first aspect that is immediately apparent in  Fig. \ref{fig:pcc_bar} is the strong (and statistically significant) anti-correlation between \sfrratio ratio and the galaxy size, even whilst controlling for stellar mass and redshift. Due to this, we can say that there is a strong anti-correlation between \sfrratio ratio and galaxy size that is not just caused by stellar mass or redshift. This implies that trends seen in previous sections are robust. We also see a statistically significant anti-correlation with redshift as expected both based on our previous results and theory \citep[e.g.][]{Mo1998}.
We see a slightly positive correlation with mass as would be expected by the size-mass relation, although this is not found to be statistically significant.
This makes it even more compelling that the relationship between size and \sfrratio is found to be statistically significant. This strongly suggests that this relationship is tracing an underlying physical relation between these two quantities.

\subsection{Stacking bursting and declining samples}

Based on the results of the previous sections, we have seen that there exists a strong anti-correlation between half-light radius and \sfrratio, which remains even after controlling for both stellar mass and redshift. This means that galaxies that are undergoing a burst  appear to be systematically more compact. 

\begin{figure*}
    \centering
    \includegraphics[width=0.99\linewidth]{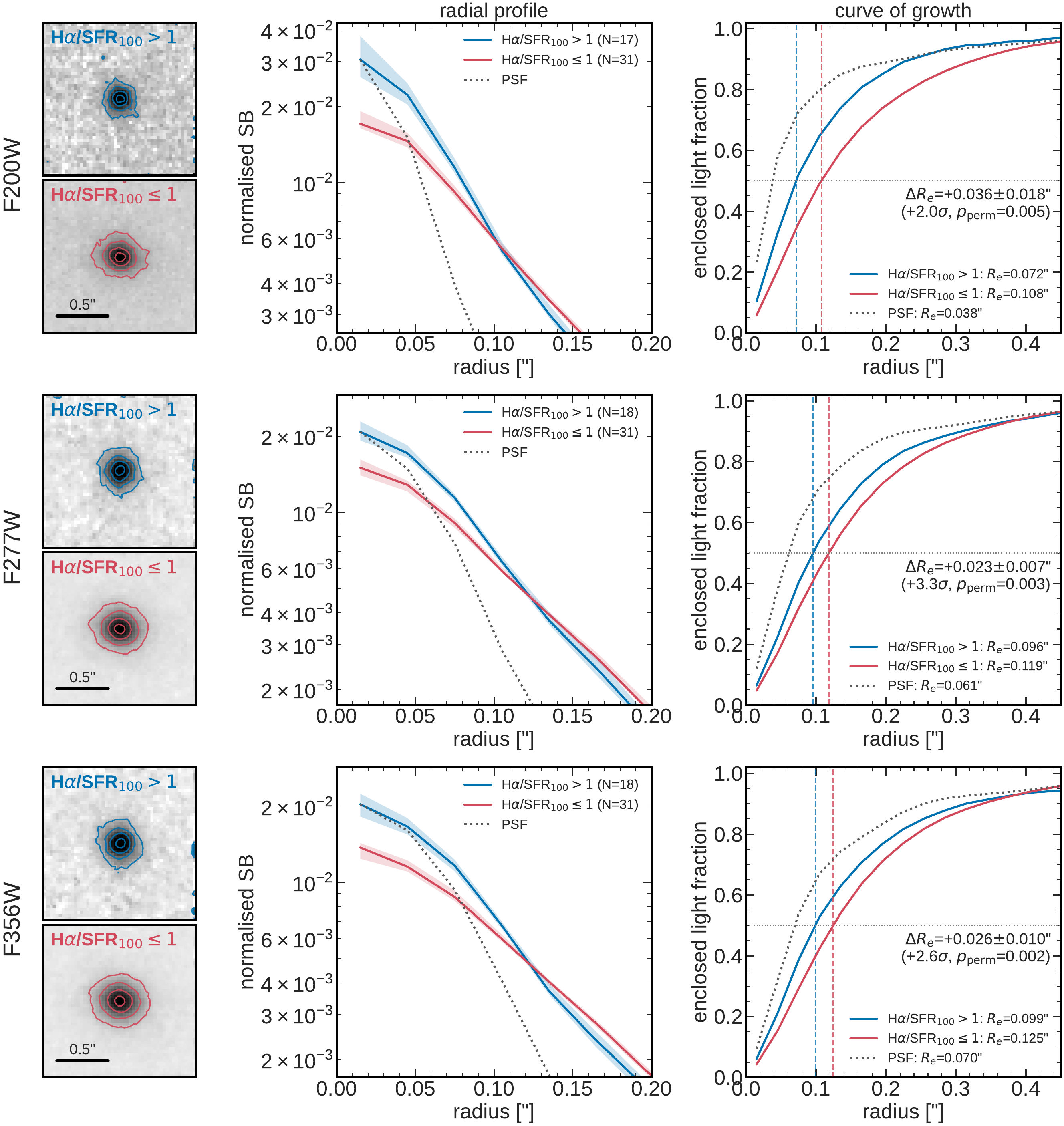}
    \caption{Left: stacks of the OASIS sample galaxies divided into those with rising or falling SFHs in F200W, F277W and F356W. Middle: normalised surface brightness profiles of the stacks with the mPSF \citep{Ji2024b} overplotted. Right: Curve-of-growth for the enclosed light of each of the stacks with the PSF and half-light radius overplotted. The figure shows that the rising stack has more centrally concentrated light than the falling stack, consistent with our interpretation of central starbursts driving galaxy half-light size.}
    \label{fig:light_profiles}
\end{figure*}

We can explore this in more detail by incorporating a stacking analysis of these two samples. As expected, the usual caveats of any stacking procedure apply; however, in this case we are using this as an independent check on our main findings obtained using \forcepho modelling. 
We use two sub-samples of galaxies corresponding to rising and falling SFHs, i.e. \sfrratio$>0$ and $<0$ respectively. We stack the imaging data for each sample in F200W, F277W, and F356W, using a median stack. We use \textsc{SEP} \citep{Barbary2016, Bertin1996} to mask out neighbouring sources and exclude them from the stack.
The resulting image for each sample is shown in the left panel of Fig. \ref{fig:light_profiles}.
We then measure the normalised surface brightness profile of the stack which is shown in the middle panel of Fig. \ref{fig:light_profiles} \citep[for more details on this procedure see][]{Baker2025} and the curve of growth of the light distribution given in the right panel. We overplot the model PSFs from \cite{Ji2024b}.
We do note that in this case we lose much of our sub-pixel resolution benefits from the \forcepho fitting, in addition to losing the ability to probe individual galaxies. We also note the caveat that the rising and falling samples will necessarily have different average stellar masses, but that in our total sample the partial correlation coefficient analysis does control for the effects of mass and finds no significant dependence of mass on sizes at fixed burstiness and redshift. Weighting the stacks to a common stellar mass and redshift distribution retains $\approx 80$\% of the size offset, albeit at reduced significance owing to the smaller effective sample.

The key aspect of this analysis is that it shows that the bursting sample are systematically more compact, i.e. the light distribution is centrally dominated, in comparison to the falling sample. This is particularly clearly seen in the radial profile and curve of growth figures, where the rising sample dominates in the centre, whilst the falling sample is more spatially extended. This is in agreement with our primary finding based on the \forcepho analysis. Similar to that analysis the same caveats apply to this; larger representative spectroscopic samples will be required to probe the population more robustly.

\subsection{The size-mass relation}

\begin{table}
\centering
\caption{Least-squares fits $\log(R_\mathrm{e}/\mathrm{kpc}) = a + b\,[\log(M_\star/M_\odot)-8.5] + c\,(z-6) + d\,B$, with $B\equiv\log(\mathrm{SFR_{H\alpha}/SFR_{100}})$, for the 46 galaxies with H$\alpha$/H$\beta$ detections. Errors from 2000 bootstrap resamplings of the sample. $\sigma$ is the rms scatter about the fit; $\Delta\sigma^2/\sigma^2$ and $p_F$ give the fractional reduction in variance and the nested $F$-test probability relative to the model with one fewer term ($M_\star$ for the second and third columns, $M_\star,z$ for the last).}
\label{tab:sizemass}
\footnotesize
\setlength{\tabcolsep}{2pt}
\begin{tabular}{@{}lcccc@{}}
\hline
 & $M_\star$ & $M_\star,z$ & $M_\star,B$ & $M_\star,z,B$ \\
\hline
$a$ & $-0.54\pm0.04$ & $-0.58\pm0.04$ & $-0.59\pm0.04$ & $-0.61\pm0.03$ \\
$b$ ($M_\star$) & $0.22\pm0.07$ & $0.22\pm0.06$ & $0.05\pm0.08$ & $0.09\pm0.07$ \\
$c$ ($z$) & -- & $-0.11\pm0.03$ & -- & $-0.08\pm0.03$ \\
$d$ ($B$) & -- & -- & $-0.42\pm0.11$ & $-0.32\pm0.11$ \\
$\sigma$ [dex] & 0.261 & 0.223 & 0.218 & 0.199 \\
$\Delta\sigma^2/\sigma^2$ & -- & 27\% & 30\% & 20\% \\
$p_F$ & -- & $<$0.001 & $<$0.001 & 0.001 \\
\hline
\end{tabular}
\end{table}

\begin{figure}
    \centering
    \includegraphics[width=0.9\linewidth]{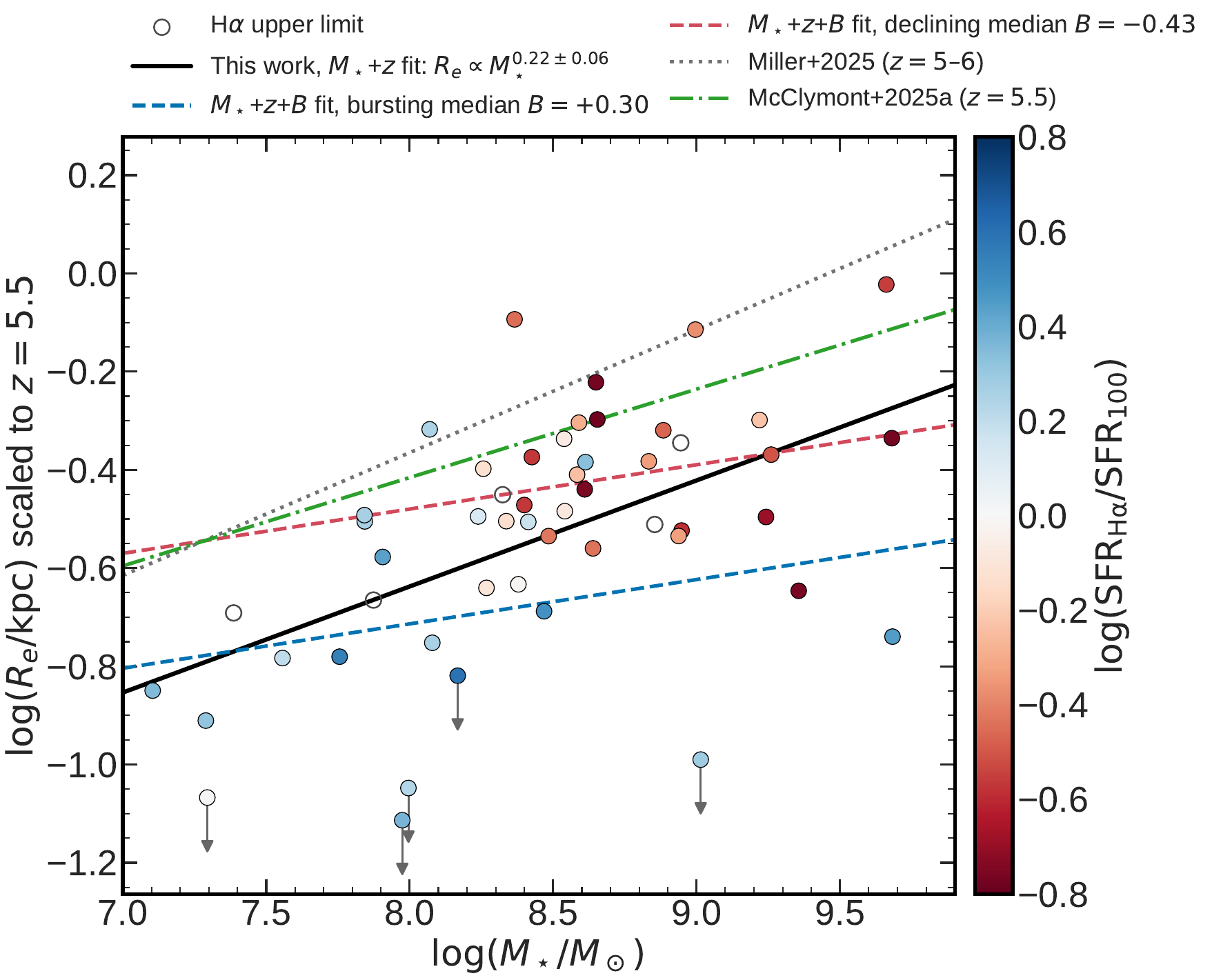}
    \caption{Effective half-light radius vs stellar mass, colour coded by burstiness (i.e. the size-mass relation). The black line gives our best-fit size-mass-redshift relation for the whole sample scaled to z=5.5, whilst blue and red dashed lines show this best fit model evaluated at the median burstiness for the rising and falling samples respectively. The size mass relations from \cite{McClymont2025} and \cite{Miller2025} are overplotted.}
    \label{fig:size-mass}
\end{figure}

Despite showing that the key drivers are burstiness and redshift, it is still important to understand the galaxies' position on the classical size-mass relation. This relation describes the positive correlation seen between the stellar mass and radius of galaxies both in the local Universe and at high-z \citep[e.g.][]{vanderWel2014, Miller2026}. Up to $z\sim3$ the size-mass relation is clearly found with strong differences between the star-forming and quiescent galaxy populations \citep{vanderWel2014}, corresponding to differences in their assembly histories, and often attributed to the effects of mergers \citep{vanDokkum2015} and/or spiral-arm modulated size evolution \citep{vanderWel2025}. 
Recent JWST works have explored this size-mass relation finding evidence for it in lower mass star-forming galaxies at high-z \citep[e.g.][]{Miller2025, Allen2025, Danhaive2026, Miller2026}, and in the quiescent galaxy population \citep[e.g.][]{Ji2024,Ito2024,Baker2025a, Kawinwanichakij2025, Miller2026}. 
Simulations vary as to whether they can produce the relation at high-z with some finding clear evidence \citep{McClymont2025} and others finding an inverted relation \citep[e.g.][which they attribute to central dust concentrations]{Roper2022, Ceverino2026}.

We explore, with many caveats, what we would see as the size-mass relation based on our sample. 
This is shown in Fig. \ref{fig:size-mass} where blue points correspond to galaxies undergoing a burst and red points correspond to those undergoing a decline. 
We also overplot the optical size-mass relation of \citet{McClymont2025} as the majority of the bands used in our \forcepho fitting correspond to rest-frame optical wavelengths, as well as that of \cite{Miller2025}. We note that they report the half-light radius as that of the semi-major axis compared to our \textsc{forcepho} circularised half-light radius which accounts for part of the size offset. 
We measure our size-mass relation by fitting a parametric form characterised by the equation
\begin{equation}
    \log(R_\mathrm{e}/\mathrm{kpc}) = a + b\,[\log(M_\star/M_\odot)-8.5] + c\,(z-6) + d\,B
\end{equation}
where $B\equiv\log(\mathrm{SFR_{H\alpha}/SFR_{100}})$.
In this way we can see how the parametrisation and scatter changes as we include more important parameters into the relation. We fit this equation using Ordinary Least Squares in 4 separate cases, the first is where $c$ and $d$ are set to zero which gives us a classical mass-size relation (corresponding to the solid black line in Fig \ref{fig:size-mass}), and then a selection of cases where we fit different combinations to explore how the scatter in the relation decreases as we include burstiness ($B$) and redshift.
The results of these fits are included in Table \ref{tab:sizemass}. The key result is that the scatter in the relation is reduced significantly by around 0.04dex by including burstiness or redshift. Including both of them reduces the scatter by 0.06dex. This is a fractional reduction in variance of 27-30\% for either term and 42\% for both. This is yet another way of seeing the importance of these two quantities in determining the size of the galaxies.

We can also use our fits to explore the redshift evolution of galaxy sizes whilst keeping mass fixed. Using the parametrisation of \citet{Danhaive2026}, given by $\log R_\mathrm{e}\propto
\gamma\log(1+z)$, our fit gives $\gamma=-1.8\pm0.4$, which is steeper than the expected evolution of the halo virial radii at a given fixed halo mass  $R_\mathrm{vir}\propto M_\mathrm{h}^{1/3}H(z)^{-2/3}\propto(1+z)^{-1}$ at
$z\gtrsim3$ \citep{Fall1980,Mo1998,Ferguson2004}. However, there is a connection between burstiness and redshift within our sample ($\rho_{B,z|M_\star}=0.36$, $p=0.014$) so part of this evolution simply reflects the higher fraction of bursty galaxies at higher redshift within our sample. If we include burstiness in the fit we reduce the redshift dependence to $\gamma=-1.3\pm0.4$ (or $R_\mathrm{e}\propto
H(z)^{-0.85\pm0.26}$) which makes it consistent with the expected growth of dark matter haloes ($\gamma=-1$, or $H(z)^{-2/3}$) \citep{Mo1998}, assuming a weakly evolving stellar-to-halo-mass ratio \citep{Behroozi2019} and an approximately constant $R_\mathrm{e}/R_\mathrm{vir}$ \citep{Kravtsov2013}. 

% Considering a given halo of fixed mass $R_\mathrm{vir}\propto M_\mathrm{h}^{1/3}H(z)^{-2/3}$ whilst at a fixed velocity $R_\mathrm{vir}\propto H(z)^{-1}$
% \citep{Mo1998,Ferguson2004}, where at $z>3$ these correspond to $(1+z)^{-1}$, and $(1+z)^{-1.5}$. 

We note that the declining galaxies alone would give $\gamma=-0.8\pm0.5$. Hence, once we compare galaxies with similar burstiness (i.e. those at the same part of their cycle), the size evolution appears to track the expected halo growth.

Another way to visualise the connection between these quantities is as a plane in 3D space. This is similar to how one would traditionally envisage a relation like the fundamental metallicity relation \citep{Mannucci2010, Curti2020, Baker2023b}. 
\begin{figure}
    \centering
    \includegraphics[width=1\linewidth]{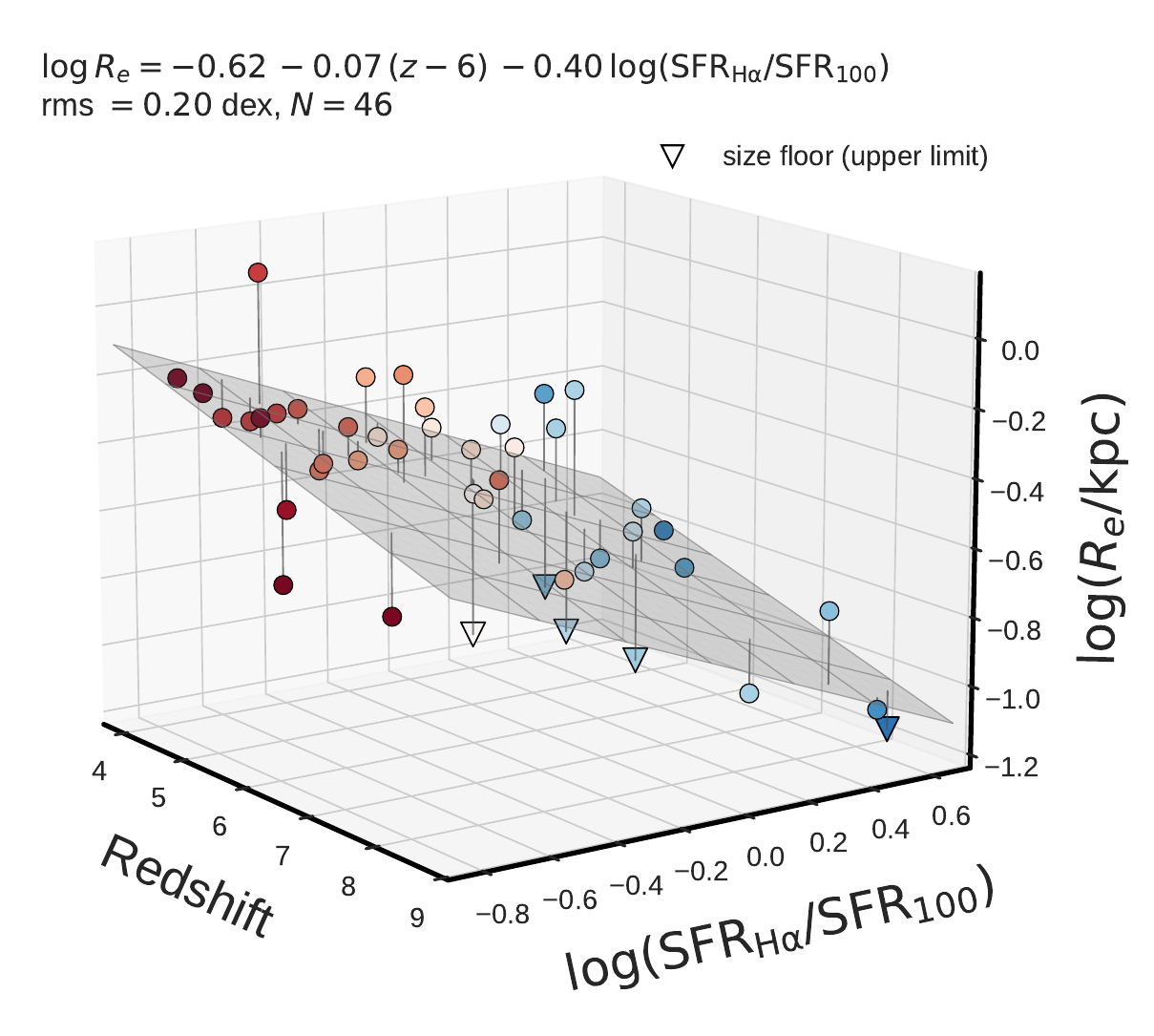}
    \caption{Size, \sfrratio, and redshift for the OASIS sample on a 3D axis (colour-coded by \sfrratio). We fit a 2D plane to the resulting relation. We see how we reduce the scatter to around 0.2dex compared to the 0.26\,dex scatter of the mass--size relation.}
    \label{fig:plane}
\end{figure}
In this case the plane is between size, burstiness and redshift. We do not have enough galaxies to explore more complicated parametric versions of a 2D surface \citep[e.g.][]{Curti2020,Baker2023b}, so we restrict our fit to a simple 2D plane to provide a visual representation of our idea. This is shown in Fig. \ref{fig:plane}.
The fit gives 
\begin{equation}
\begin{split}
\log\left(\frac{R_\mathrm{e}}{\mathrm{kpc}}\right) =
(-0.62\pm0.03) + (-0.072\pm0.025)\,(z-6)
\\ + (-0.40\pm0.09)\,\log\left(\frac{\mathrm{SFR_{H\alpha}}}{\mathrm{SFR_{100}}}\right),
\end{split}
\label{eq:plane}
\end{equation}
with a scatter of just $\sigma=0.20$dex.  Uncertainties are from bootstrap resampling 2000 times. We can see that the scatter of 0.2 dex is significantly less than that of the mass-size relation for the sample $\sigma_{MS}=0.261$dex.
The scatter for the 2D plane fit matches the scatter of the size-mass relation when we included redshift and burstiness in our fitting, providing a complementary view that almost the entire relationship can be described by a more fundamental relationship between size, burstiness and redshift.

\subsection{Secondary relation between ionisation parameter and size}
\label{sec:U}

\begin{figure*}
    \centering
    \includegraphics[width=0.49\linewidth]{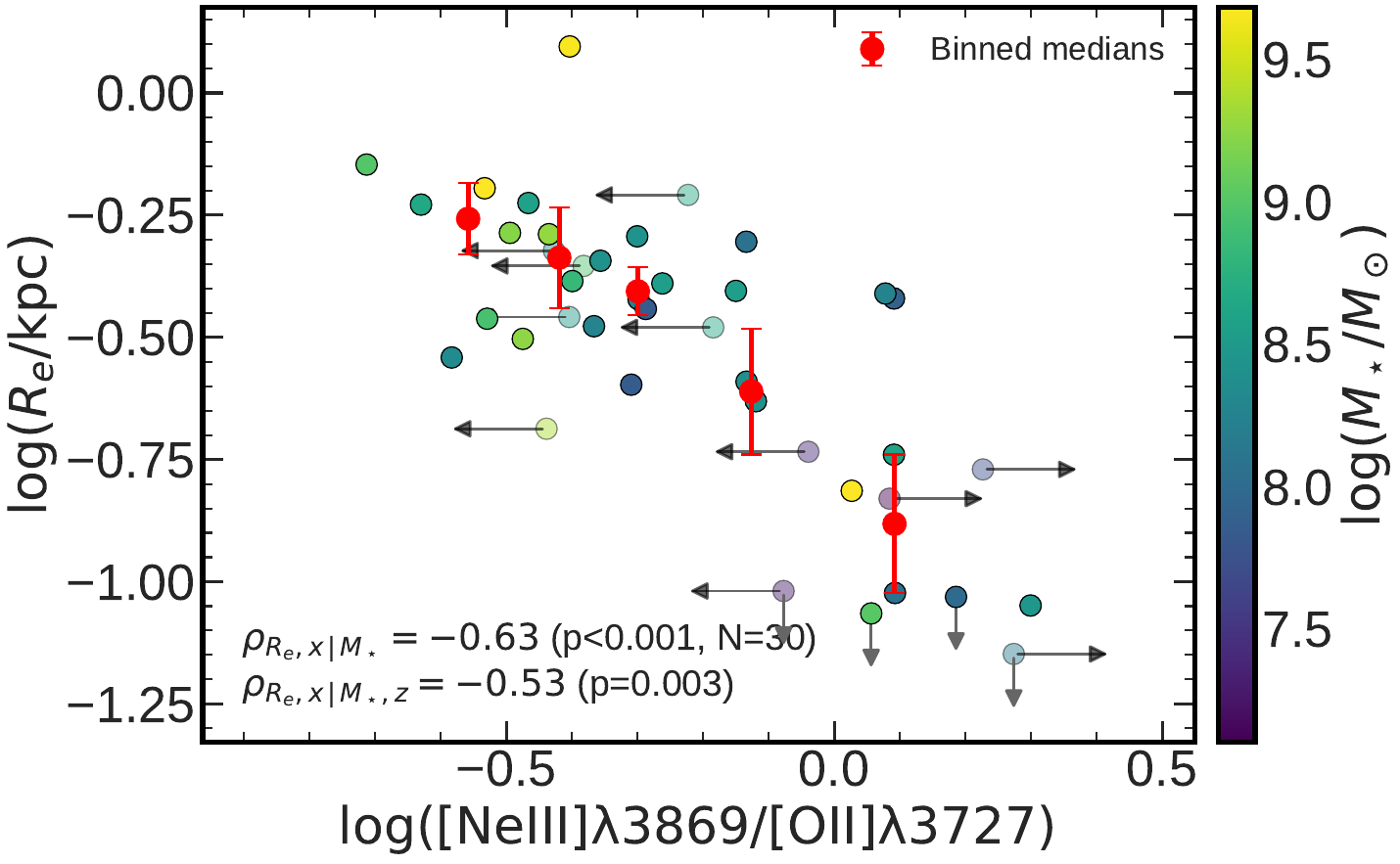}
    \includegraphics[width=0.49\linewidth]{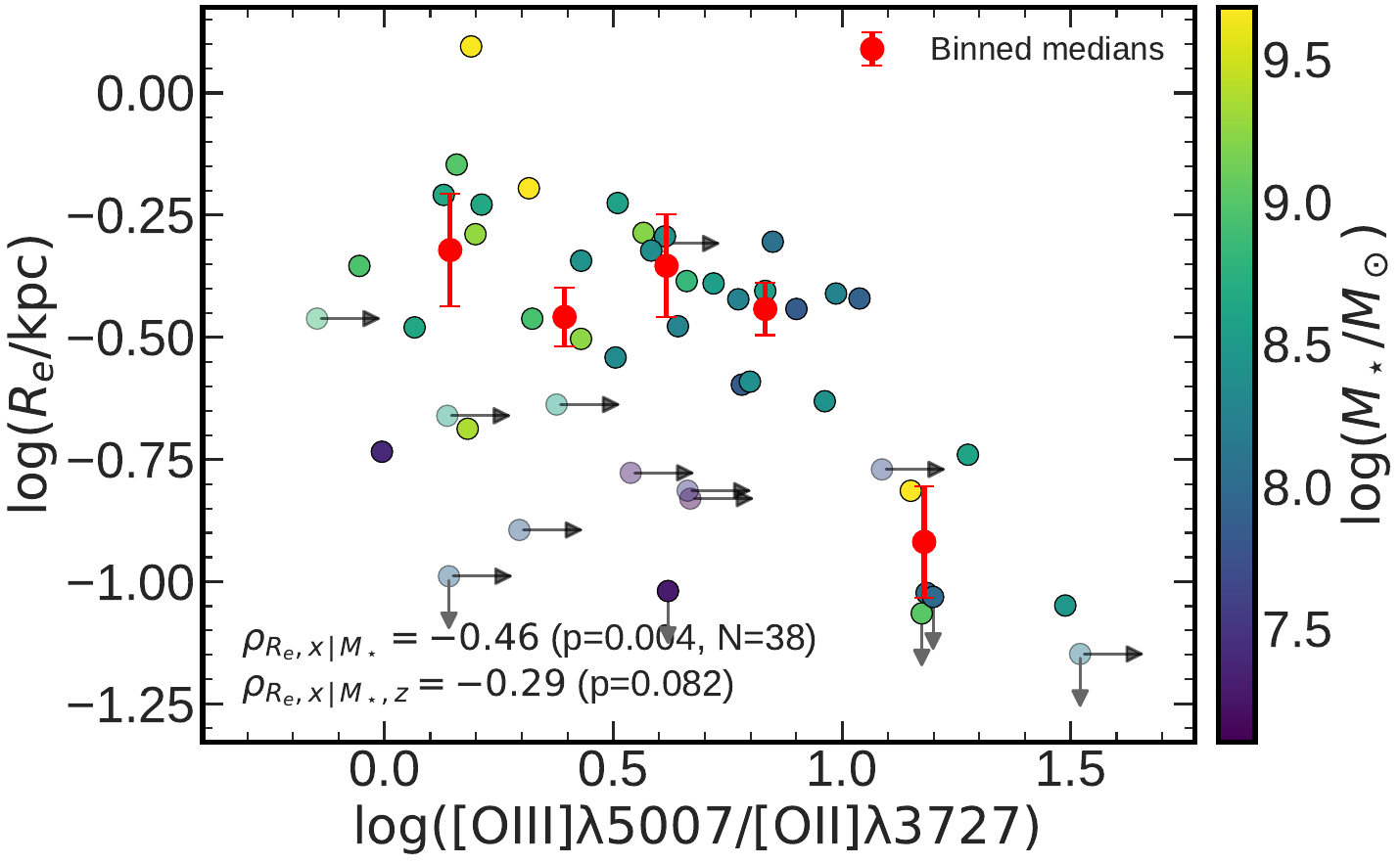}
    \caption{Effective half-light radius vs log(\neiiiL3869/\oiiL3727,29) (left) and log(\oiiiL5007/\oiiL3727,29) (right), colour-coded by stellar mass. Both line ratios are strong tracers of the ionisation parameter, and both show a strong anti-correlation with galaxy size. This indicates that galaxies with higher ionisation parameter (consistent with those undergoing a burst of star-formation) are systematically more compact.}
    \label{fig:size-U}
\end{figure*}

If bursty galaxies really are more compact than declining galaxies we would also expect to see correlations between other properties that correlate with burstiness and size. This need not be intrinsic, rather a by-product of their underlying relation with burstiness \citep[see e.g. ][as notable examples]{Baker2022, Baker2023}. In OASIS Paper~II we show a clear correlation between \sfrratio and both 
$\NeOtwo\equiv\log(\mathrm{[Ne\,\textsc{iii}]}\lambda3869/\mathrm{[O\,\textsc{ii}]}\lambda\lambda3727,29)$
% log(\neiiiL3869/\oiiL3727,29) and log(\oiiiL5007/\oiiL3727,29), 
and $\Othreetwo\equiv\log(\mathrm{[O\,\textsc{iii}]}\lambda5007/\mathrm{[O\,\textsc{ii}]}\lambda\lambda3727,29)$
themselves both strong tracers of the ionisation parameter (U) within galaxies. This can be understood as being driven by burstiness, galaxies that are undergoing a burst of star-formation produce massive OB type stars that produce the strongest ionisation, whilst galaxies with declining SFRs have fewer massive stars (if any) available to ionise. This would result in a strong link between burstiness and these tracers of the ionisation parameter. Therefore, we should also see a strong relation between size and these ionisation parameter diagnostics. 

In Fig. \ref{fig:size-U} we show galaxy size vs Ne3O2 (left) and also O32 (right), colour-coded by stellar mass. We again see a strong anti-correlation with galaxies with higher ionisation parameter being more compact. However, this is simply an effect of the underlying relationship between burstiness and size. We clarify this by again measuring \pccs between the two quantities, whilst controlling for burstiness, in this case the relationship is no longer statistically significant. 

Another alternative is that, not only are Ne3O2 and O32 tracers of the ionisation parameter, they also can trace metallicity \citep[e.g.][]{Maiolino2019}. We briefly test whether the relationship is seen via using the code \textsc{Genesis-metallicity} \citep{Langeroodi2026} which uses a non-parametric approach to determine metallicities given any available line diagnostics. 
This necessitates a significant reduction in our sample size to 36 galaxies.
We find a strong positive correlation between metallicity and size at fixed stellar
mass and redshift ($\rho_{R_\mathrm{e},Z|M_\star,z}=0.40$, $p=0.018$), i.e.\ more compact
galaxies are more metal poor, and the bursting galaxies are also systematically more
metal poor (median $12+\log(\mathrm{O/H})=7.71$ compared with 8.20 for the declining
galaxies). This is in agreement with results from the literature
\citep[e.g.][]{Langeroodi2023}, but once burstiness is controlled for the
size--metallicity relation is no longer significant
($\rho_{R_\mathrm{e},Z|M_\star,z,B}=0.25$, $p=0.16$), whereas the size--burstiness
relation remains significant when controlling for metallicity
($\rho_{R_\mathrm{e},B|M_\star,z,Z}=-0.42$, $p=0.014$).
Conversely, the higher ionizing photon production efficiency of the more metal-poor, compact galaxies could bias \sfrratio upwards; however, the H$\alpha$ conversion we adopt is already calibrated for metal-poor populations (Sect.~\ref{sec:data}), so only the smaller metallicity differences within our sample contribute.

Overall this is again reconfirming that the key relation appears to be between burstiness and size with an additional dependence on redshift. The other relations that appear seem to stem from these two more fundamental relations.

\section{Discussion}

\subsection{What governs size evolution and burstiness?}

\label{d.s.ss.split_into_rising_and_falling}
We have shown that there exists a strong link between \sfrratio and the half-light effective radii for galaxies in the OASIS sample, a sample that should be representative of the overall galaxy populations at these redshifts.
We have shown that this anti-correlation remains, even after controlling for further dependencies on stellar mass and redshift and therefore likely corresponds to an underlying physical mechanism. The primary drivers of galaxy size for the OASIS sample are both \sfrratio and redshift.

In this section, we discuss potential mechanisms in more detail. The simplest dependence to explain is that of redshift as it has been expected and shown that higher redshift galaxies should be intrinsically smaller, with their sizes growing over time due to the growth of the underlying dark matter haloes \citep{Fall1980, Mo1998, Ferguson2004}. Therefore our finding of a redshift dependence is fully consistent with this.

The link with burstiness is more revealing.
As outlined in the introduction, we propose that the strong anti-correlation between size  and burstiness is driven by the effect of compaction driven central starbursts. We reiterate again that in this work we are exploring galaxy observed half-light radii, rather than half-mass radii - therefore our light distribution, and hence what we call size, is dominated by the young stellar populations. Therefore, our sizes trace where the star-formation is taking place in a galaxy. Hence, finding that galaxies undergoing a burst, as traced by \sfrratio, have smaller half-light radii, and more centrally concentrated light distributions, suggests that this is driven by central starbursts which have the effect of shrinking our half light radii.
This would serve to decrease observed galaxy sizes (as measured via their half-light radii) for galaxies undergoing a burst compared to those with declining SFRs. 

A bright central AGN would also be capable of shrinking our half-light radii in a similar way to a central starburst, but we do not see any evidence in the spectra for our sample of strong AGN contamination.
Another alternative could be inside-out quenching, which would make bursting galaxies seem smaller than lulling or mini-quenched ones. However, although the declining galaxies are larger on average (Fig.~\ref{fig:light_profiles}), their sizes do not increase with the strength of the decline, as would be expected for inside-out quenching: the anti-correlation is driven by the galaxies in burst phases (i.e. \sfrratio$>0$), whilst the declining galaxies show no clear trend (Fig.~\ref{fig:size_burstiness}).

\begin{figure*}
    \centering
    \includegraphics[width=0.89\linewidth]{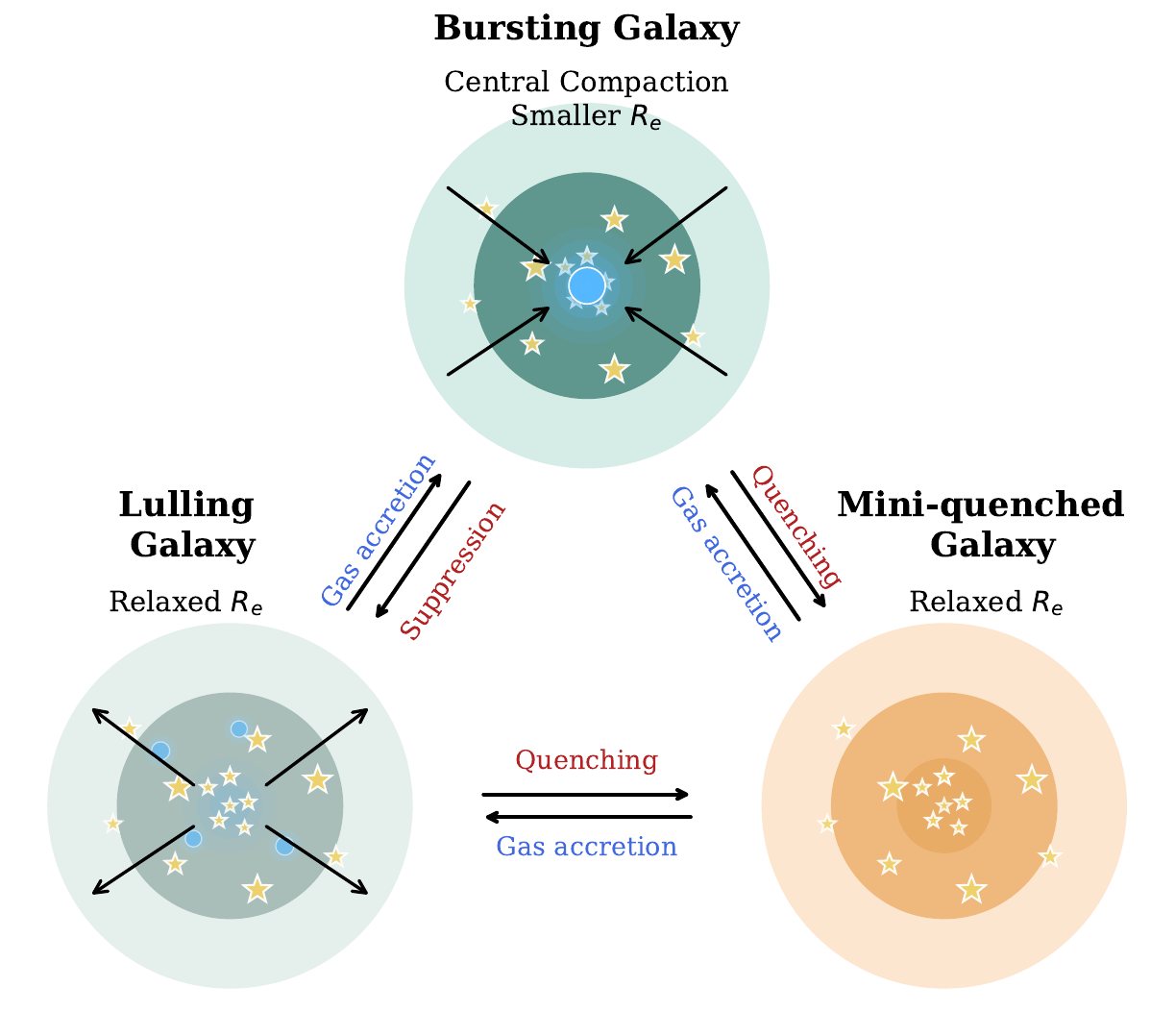}
    \caption{Schematic diagram showing our interpretation of the observed anti-correlation between burstiness and half-light radius and how it connects in with lulling and mini-quenched galaxies. The small blue circles represent active star-formation (i.e. gas or young stars) whilst the stars represent older stellar populations \citep[partly inspired by][]{McClymont2025}. The arrows indicate the direction of change of the half-light radius. It shows how a central starburst shrinks the half light radius, before a shutdown of star-formation then comparatively expands it prior to fresh gas accretion. Each of these phases can transition into each other via the processes of fresh gas accretion or feedback and gas depletion.}
    \label{fig:schematic}
\end{figure*}

Our fiducial interpretation is detailed in Fig. \ref{fig:schematic}, where we include possible mechanisms and link them to the galaxies in our sample.
If we start with a galaxy undergoing a central starburst caused by a significant gas reservoir in the central region \citep[perhaps itself caused by mergers, violent disc instabilities, or cold streams][]{Hernquist1989,Dekel2006, Zolotov2015} this has the effect of shrinking the half-light radii (upper panel). The large central gas reservoir results in rapid star-formation in this central region. 
This could then lead to either a lulling (lower left) or mini-quenched (lower right) galaxy, based on the resulting feedback/gas depletion. 

The first of these is that the effect of stellar feedback (or potentially AGN feedback) or even just gas starvation could then lead to a shutdown of star-formation within the central region, with weak residual star formation remaining (lower left panel). This corresponds to a lulling-like galaxy, i.e. a visible Balmer break and weak emission lines corresponding to residual star formation. Even if the star-formation does not continue in the outskirts this can still have the effect of increasing the half-light radius due to a relaxation of the light profile, i.e. it will no longer remain centrally concentrated. 
The second is similar, whereby feedback may entirely shut off star-formation within the galaxy leading to a mini-quenched galaxy with no sign of active star-formation (lower right). This would also have the effect of increasing the half-light radius of the galaxy, again due to a relaxation of the overall light profile (i.e. it no longer being dominated by a central starburst). 
This relaxation effect essentially could correspond to either of the lower panels.

Either of the states (lulling or mini-quenched) has the effect of increasing the half light-radius as the ratio between the luminosity in the centre to the outskirts decreases as there is no longer the central starburst. This essentially is also another form of outshining, whereby the light from the young stars dominates the galaxy light profile.
Potentially there can also be a transition between the lulling (weakly star-forming) and mini-quenched (non-star-forming) states caused by minor gas accretion or further quenching. The impact of this on sizes would likely be small, but would also depend on whereabouts any gas accretion/star-formation is/was taking place within the galaxy.

These effects are clearly found in some simulations, with large stochasticity caused by the associated interplay between star formation, gas accretion, and feedback \citep[e.g.][]{Zolotov2015,El-Badry2016, Tacchella2016b, McClymont2025}. 
However, our results are in disagreement with other simulations such as \textsc{FLARES} \citep{Roper2022} and \textsc{FirstLight} \citep{Ceverino2026}. As shown in Appendix Sec. \ref{r.ss.dust} we find no evidence for dust trends being responsible for the size-burstiness relation and hence find greater agreement with the results of the \textsc{thesan-zoom} simulations, which is likely due to the inclusion of bursty star formation driven by the multiphase ISM prescriptions within \textsc{thesan-zoom} \citep{McClymont2025}.

\subsection{Compaction and Central Starbursts}

We show clear evidence for a negative relationship between burstiness and half-light radius which we believe to be caused by central starbursts. 
The key requirement for a central starburst is a large excess of gas within the central region of a galaxy. There are multiple potential mechanisms for this process, such as, violent disc instabilities in which the disc becomes unstable, fragments into clumps which migrate to the centre of the galaxy causing compaction and driving a central starburst \citep[e.g.][]{Barro2013,Dekel2014}; `wet' (i.e. gas rich) mergers \citep{Hernquist1989}, which could funnel gas to the central region, or potentially counter rotating streams feeding the central region \citep[][]{Zolotov2015}. 
These scenarios have been heavily studied at lower-redshift and around cosmic noon and fit into a framework in which galaxies first build dense cores via compaction, followed by inside-out growth of their discs and inside-out quenching \citep[e.g.][]{vanDokkum2010,Tacchella2015, Nelson2016, Ji2024b}. In the local Universe, while we generally observe inside-out growth in discs \citep[e.g.,][]{Pezzulli2016}, occasional outside-in growth (or shrinking) is also observed in dwarf galaxies \citep{Zhang2012}, i.e. precisely those gas-rich local galaxies that may best resemble early systems observed by OASIS.

The importance of these effects at higher redshift is still subject to debate. Compaction and central starbursts are found in simulations in \citet{McClymont2025} using the \textsc{thesan-zoom}, with observational evidence coming from FUV rest-frame sizes appearing smaller as galaxies rise above the MS in \citet{Danhaive2026}. 

The evidence for inside-out growth on the other hand is still debated. Using grism spectroscopy \citet{Matharu2024} find that H$\alpha$ sizes are extended, traditionally a sign of inside-out growth, whilst \citet{Danhaive2026} find that NUV, FUV and optical sizes are broadly consistent, which suggests a lack of inside-out growth.
Part of the debate is how well H$\alpha$ sizes actually trace star-formation with suggestions that it may rather trace more diffuse ionised gas that is further from the actual active star-formation \citep[][]{McClymont2025}.
% but observationally it remains unclear. \cite{Matharu2024} find evidence for inside-out growth based on H$\alpha$ sizes; however, some other observational studies do not seem to find evidence for this \citep{Danhaive2025b} and it remains unclear how well H$\alpha$ emission actually traces star-formation radii as opposed to the more diffuse ionised gas located further out from the sites of active star formation  \citep{McClymont2025}. 
Potentially other effects such as strong outflows can also affect galaxy morphology in medium band filters \citep[e.g.][]{Zhu2025}, but is not an issue here due to our multi wide band coverage. Using just the bands free of strong emission lines, the size–burstiness anti-correlation remains statistically significant.
% Individual galaxies growing inside-out via central cores have been observed up to z$\sim$7.5, such as the example in \citet{Baker2025}, but the behaviour of populations remains unclear. 

Our results are consistent with a form of growth where galaxies that are undergoing a burst (i.e. rising SFHs) are more compact than those declining (falling SFHs).
This leads us to favour central starbursts as the explanation for this.

\subsection{Impact on metallicity gradients}
In principle the central starburst should be traceable via stellar population and gas  gradients such as metallicity and/or age gradients. Galaxies undergoing gas-rich central starbursts should have evidence for metal dilution in the centre, due to dilution of metal poor gas. Similarly, the stellar population age should also be affected, with the younger stellar populations tracing the sites of most active star-formation.
This should lead to correlations between metallicities and size as we showed in Sec. \ref{sec:U} and also found in \cite{Langeroodi2023} where the more offset a galaxy is from the FMR, i.e. how out of equilibrium it is due to processes such as central starbursts, the smaller its respective half-light radius.
Similar results are also hinted at in \cite{Tripodi2024}.

However, metallicity gradients are affected by a complex combination of different effects, including gas accretion, star-formation, winds, turbulence, and radial transport, in addition to the strong systematics involved in their accurate observational measurement, and this makes them complicated to interpret \citep{Maiolino2019}. Most high-z works have found flat metallicity gradients \citep[e.g.][]{Koller2026, Tripodi2024}, although others have found metal enhanced centres \citep{Li2025}. Simulations show that bursty star-formation may lead to flat metallicity gradients caused by increased turbulence \citep{Garcia2026}. 

The effect on age gradients is also complicated to measure. Simulations have found a variety of predictions, often involving significant amounts of mixing of different stellar population ages and non-trivial age gradients \citep[e.g.][]{El-Badry2016, Dome2024}. More spatially resolved NIRSpec IFS studies will be required to test these effects in the future.

\subsection{Implications for galaxy population studies}

The uncovered relation between galaxy size and burstiness has important implications for our studies of high-redshift galaxy populations. Galaxies that are undergoing central starbursts are more compact. This has ramifications for our understanding of completeness, because often population-wide studies simply employ completeness tests based on observed luminosity and hence mass-to-light ratios, without necessarily taking into account galaxy sizes.
In these cases, any magnitude or flux limited survey would have an over representation of galaxies undergoing bursts, not only because they are inherently more luminous \emph{but also because they are more compact.}
This would skew population based studies towards bursting systems, beyond simple considerations of their greater luminosity, and motivates including size measurements in completeness studies.

\begin{figure}
    \centering
    \includegraphics[width=0.99\linewidth]{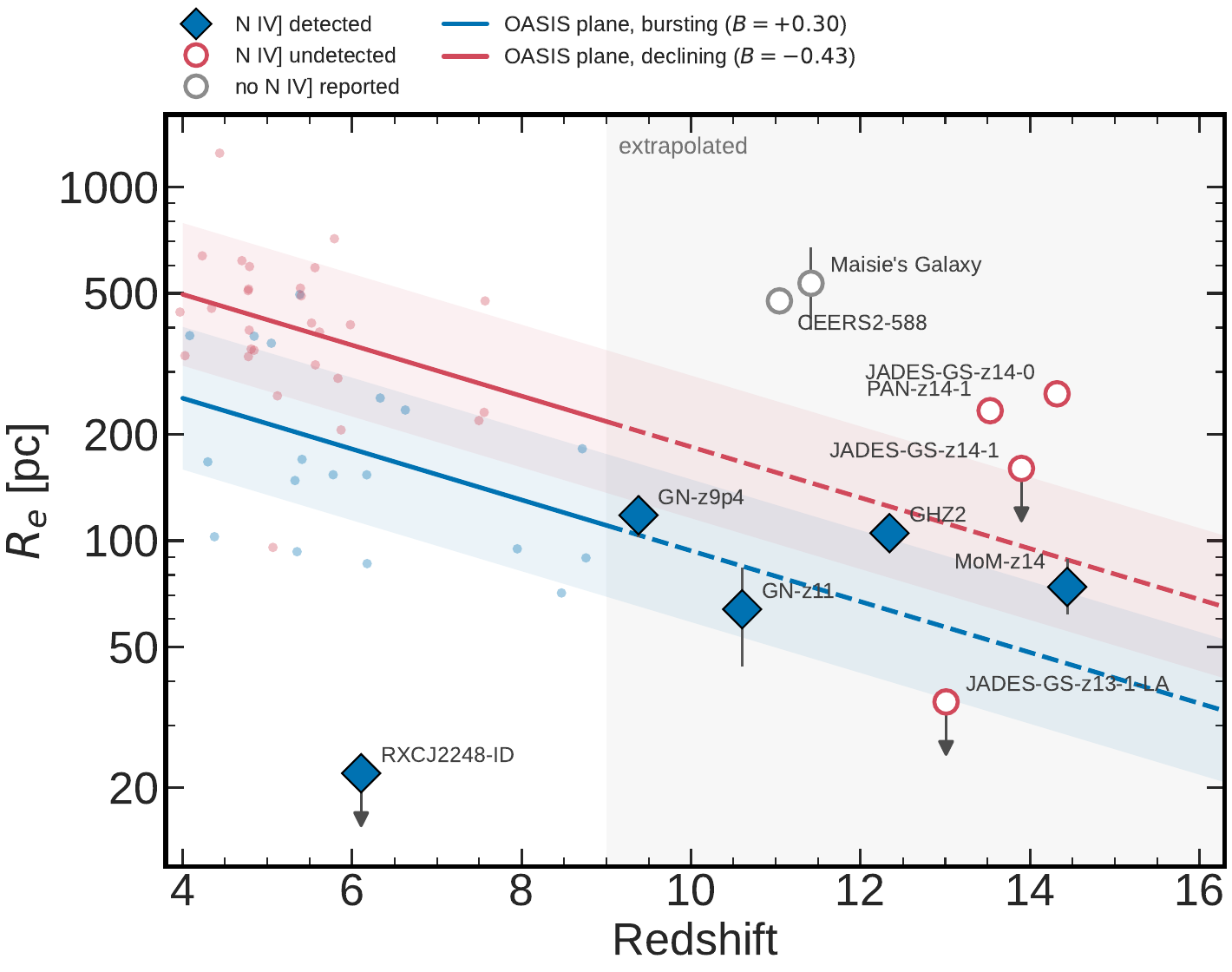}
    \caption{Effective half-light radius vs redshift. Inspired by the figure of \cite{Donnan2026}, we separate out $z>10$ galaxies into NIV] detected or undetected. We show that extrapolating the behaviour of the OASIS sample to high redshift would predict a size difference between the galaxies in a burst phase, as traced by NIV] (amongst other properties), and those undergoing a decline, whereby those undergoing a burst are more compact.}
    \label{fig:niv}
\end{figure}

\subsection{Potential link to UV bright galaxies}
Recent JWST results have suggested a potential bimodality in the size distributions of the $z>10$ galaxy population \citep[e.g.][]{Harikane2025b}, with some being extremely compact \citep{Tacchella2023, Bunker2023, Castellano2024, Naidu2025} and others being more spatially extended \citep{ArrabalHaro2023,Donnan2026, Carniani2024, Harikane2026}. Although, the sample size at $z>10$ is much smaller and galaxy sizes have been measured in a variety of different ways, it is tempting to speculate that perhaps the same processes of central starbursts (e.g. Fig. \ref{fig:schematic}) are happening at these earliest of times. 

\cite{Harikane2025b} and later \cite{Naidu2025} highlighted that compact galaxies have higher 
\NIV\
 equivalent widths than more extended galaxies. This diagnostic (amongst other things) is a tracer of higher ionisation parameter, as \NIV\ requires a hard, intense radiation field from either young massive stars or an AGN. This line also requires high N/O and N/C abundances \citep[e.g.][]{Cameron2023b, Schaerer2024} which has also often been connected to dense, compact star-formation such as massive star clusters \citep[e.g.][]{Senchyna2024}. This links the strong \NIV\ emission to both ionisation conditions and to compactness and central starbursts.
 \cite{Roberts-Borsani2026} explored the equivalent width of \CIV\ compared to galaxy sizes, linking this trend to burstiness amongst the $z>10$ population.

Therefore, this would be consistent with our findings in Sec. \ref{sec:U}, where we found that the higher the ionisation parameter (as traced by line ratios), the more compact the galaxy. Our result would point towards this being driven by burstiness caused by central starbursts. The relationship between EW(NIV]) and size would then simply be a byproduct of not being able to trace \sfrratio for these galaxies. Fig. \ref{fig:niv}, inspired by \cite{Donnan2026,Naidu2025}, shows this, based on extrapolating our plane fit (see Fig. \ref{fig:plane}) to much higher redshifts. 

It includes a compilation of galaxies from \cite{Naidu2025, Carniani2024, Donnan2026, Castellano2024, Tacchella2023, Finkelstein2022, ArrabalHaro2023, Bunker2023b, Schaerer2024, Witstok2025, Topping2025}.
Naturally, this is an approximation with many caveats, but it does appear to predict a clear average size difference between the populations which would correspond to burstiness, with those galaxies undergoing bursts, showing higher NIV] and being more compact. 
NIV] detected galaxies at $z>9$ have a median size of 90pc compared to 197pc for the undetected ones. This is 2.2 times larger for the undetected. Our extrapolation based on the OASIS galaxies would suggest that the difference should be around 2 times, which is consistent with what is observed. This suggests that burstiness is likely to contribute towards these size differences, although we note that part of the EW(NIV]) - size relation may also stem from abundance differences rather than just burstiness (although these may of course be linked to burstiness themselves).

Other scenarios are possible such as AGN, which could shrink the half-light radii in a similar way to central starbursts, but currently we do not see strong evidence for AGN contamination in either our sample or the majority of the $z>10$ galaxies from the literature \citep[although the exact nature of GNz11 is still debated, e.g.][]{Bunker2023b,Maiolino2024, CrespoGomez2026, Chen2026}. 
Clearly larger sample statistics and further mid-IR constraints from MIRI (corresponding to the rest frame optical for these galaxies) will better constrain the stellar population properties, and will be necessary to investigate this in more detail.

\section{Conclusions}

In this work we utilised the OASIS survey (OASIS Paper~I) to explore the effect of galaxy burstiness on the observed (rest UV/optical) half-light radii of a representative sample of 51 galaxies from $z\approx4-9$. Our findings are as follows:

\begin{itemize}
    \item We show a clear inverse relationship between burstiness (as measured by \sfrratio) and half-light radius for a representative sample of high-redshift galaxies ($z=4-9$) with deep JWST spectroscopy and photometry probing well below the SFMS.

    \item We employ partial correlation coefficients to isolate out potential secondary dependencies on mass and redshift, and show that, when controlling for mass and redshift, the key driver of galaxy sizes appears to be both burstiness and redshift, with galaxies undergoing a burst having reduced sizes compared to those with declining SFRs. 

    \item Consistent with this result, we find that a stacked sample of bursting galaxies  have more centrally concentrated light profiles (and hence smaller half-light radii) than the corresponding stacked sample of galaxies with declining SFRs.

    \item We show that we can obtain the mass-size relation from our sample, but that the scatter in the relation is significantly reduced by 0.06dex by including dependencies on both burstiness and redshift. We show that we obtain the same amount of scatter by excluding the stellar mass entirely from the relationship and instead fitting a 2D surface in the 3D space between burstiness, size and redshift.

    \item We show that this relationship between burstiness and size also leads to secondary relations between other quantities that correlate with burstiness, such as ionisation parameter tracers, metallicity, and size. This could explain the bimodality seen in the $z>10$ UV luminous galaxy population whereby galaxies with higher NIV] EW are more compact. Instead, our results suggest that this is driven by their increased burstiness via central starbursts.

    \item The relationship between size and burstiness strongly motivates consideration of galaxy sizes when measuring completeness in high-z surveys to avoid over-representation of centrally starbursting systems.

\end{itemize}

To explore the links between burstiness and size further, we require a significantly larger complete spectroscopic survey targeting galaxies both above and below the MS. This is crucially important for many high-z population-wide analyses and is particularly needed for these low-mass bursty galaxies.
% Whilst this will be challenging, due to the difficulty in identifying lulling or mini-quenched galaxies from photometry alone, it is crucial in accurately probing the sub-MS galaxy populations at these redshifts.
% This has strong implications for our understanding of high-redshift galaxy evolution, as due to sizes being linked to central strabursts

% \section*{Data Availability}
%  %The raw JWST data is available via the DAWN JWST Archive (\href{https://dawn-cph.github.io/dja/index.html}{DJA}). 
%  The raw JWST data is available via MAST or the DAWN JWST Archive (\href{https://dawn-cph.github.io/dja/index.html}{DJA}). 
%  The derived products are available upon reasonable request to the first authors. 

% \begin{acknowledgements}

% WMB would like to acknowledge support from DARK via the DARK Fellowship. This work was supported by a research grant (VIL54489) from VILLUM FONDEN. TJL gratefully acknowledges support from the Swiss National Science Foundation through a SNSF Mobility Fellowship and from the NASA/JWST Program OASIS (PID \#5997). FDE acknowledges support by the Science and Technology Facilities Council (STFC), by the ERC through Advanced Grant 695671 ``QUENCH'', and by the UKRI Frontier Research grant RISEandFALL.

\begin{acknowledgements}
WMB gratefully acknowledges support from DARK via the DARK Fellowship. This work was supported by a research grant (VIL54489) from VILLUM FONDEN.
TJL gratefully acknowledges support from the Swiss National Science Foundation through a SNSF Mobility Fellowship and from the NASA/JWST Program OASIS (PID \#5997).
FDE acknowledges support by the Science and Technology Facilities Council (STFC), by the ERC through Advanced Grant 695671 ``QUENCH'', and by the UKRI Frontier Research grant RISEandFALL.
AJB and JC acknowledge funding from the ``FirstGalaxies'' Advanced Grant from the European Research Council (ERC) under the European Union's Horizon 2020 research and innovation programme (Grant agreement No.~789056).
SC acknowledges support by European Union's HE ERC Starting Grant No.~101040227 -- WINGS. 
A.L.D. thanks the University of Cambridge Harding Distinguished Postgraduate Scholars Programme and Technology Facilities Council (STFC) Center for Doctoral Training (CDT) in Data intensive science at the University of Cambridge (STFC grant number 2742605) for a PhD studentship.
QD acknowledges support from a PhD studentship awarded by Trinity College, University of Cambridge.
JMH acknowledges support from the Evolving Universe Fellowship, which is made possible by a generous donation from Dr.~Keiko Miwa Ross. JMH is also supported by the JWST/NIRCam Science Team contract to the University of Arizona, NAS5-02105, and by JWST Programs 3215 and 8544.
BDJ acknowledges support from the JWST/NIRCam Science Team contract to the University of Arizona, NAS5-02105, and JWST Program 3215.
PR acknowledges support from the University of Texas at Austin Cosmic Frontier Center.
BER acknowledges support from the NIRCam Science Team contract to the University of Arizona, NAS5-02105, and JWST Program 3215.
ST acknowledges support by the Royal Society Research Grant G125142.
JAAT acknowledges support from the Simons Foundation and JWST Programs 2959 and 3215. Support for Programs 2959 and 3215 was provided by NASA through a grant from the Space Telescope Science Institute, which is operated by the Association of Universities for Research in Astronomy, Inc., under NASA contract NAS 5-03127.
This work is based on observations made with the NASA/ESA/CSA \textit{James Webb
Space Telescope}. The data were obtained from the Mikulski Archive for Space
Telescopes (MAST) at the Space Telescope Science Institute, which is operated by
the Association of Universities for Research in Astronomy, Inc., under NASA
contract NAS 5-03127 for \textit{JWST}. These observations are associated with
programmes GO 5997 (OASIS), GTO 1180, 1210 and 1286 and GO 3215 (JADES),
GO 1963 \citep[JEMS][]{Williams2023} and GO 1895 \citep[FRESCO][]{Oesch2023}. The authors acknowledge the JEMS and FRESCO
teams for developing their observing programmes with a zero-exclusive-access period
The authors acknowledge usage of Claude-Opus 5.5 for assistance with punctuation, minor rephrasing of occasional sentences, and occasional assistance with coding.
We acknowledge use of \textsc{astropy} \citep{AstropyCollaboration2013}, \textsc{numpy} \citep{harris2020array}, \textsc{ppxf} \citep{Cappellari2023}, \textsc{matplotlib} \citep{Hunter:2007}, \textsc{ForcePho} (B.~D.~Johnson et al., in prep.), and \textsc{topcat} \citep{Taylor2005}.
\end{acknowledgements}

% We acknowledge use of \textsc{astropy} \citep{AstropyCollaboration2013},  \textsc{numpy} \citep{harris2020array}, \textsc{ppxf} \citep{Cappellari2023}, 
% \textsc{matplotlib} \citep{Hunter:2007}, \textsc{ForcePho} (B. Johnson et al, in prep, and \textsc{topcat} \citep{Taylor2005}. 

% \end{acknowledgements}

\bibliographystyle{aa}
\bibliography{refs-combined}

\begin{appendix}

\section{The effect of dust} 
\label{r.ss.dust}

\begin{figure}
    \centering
    \includegraphics[width=1\linewidth]{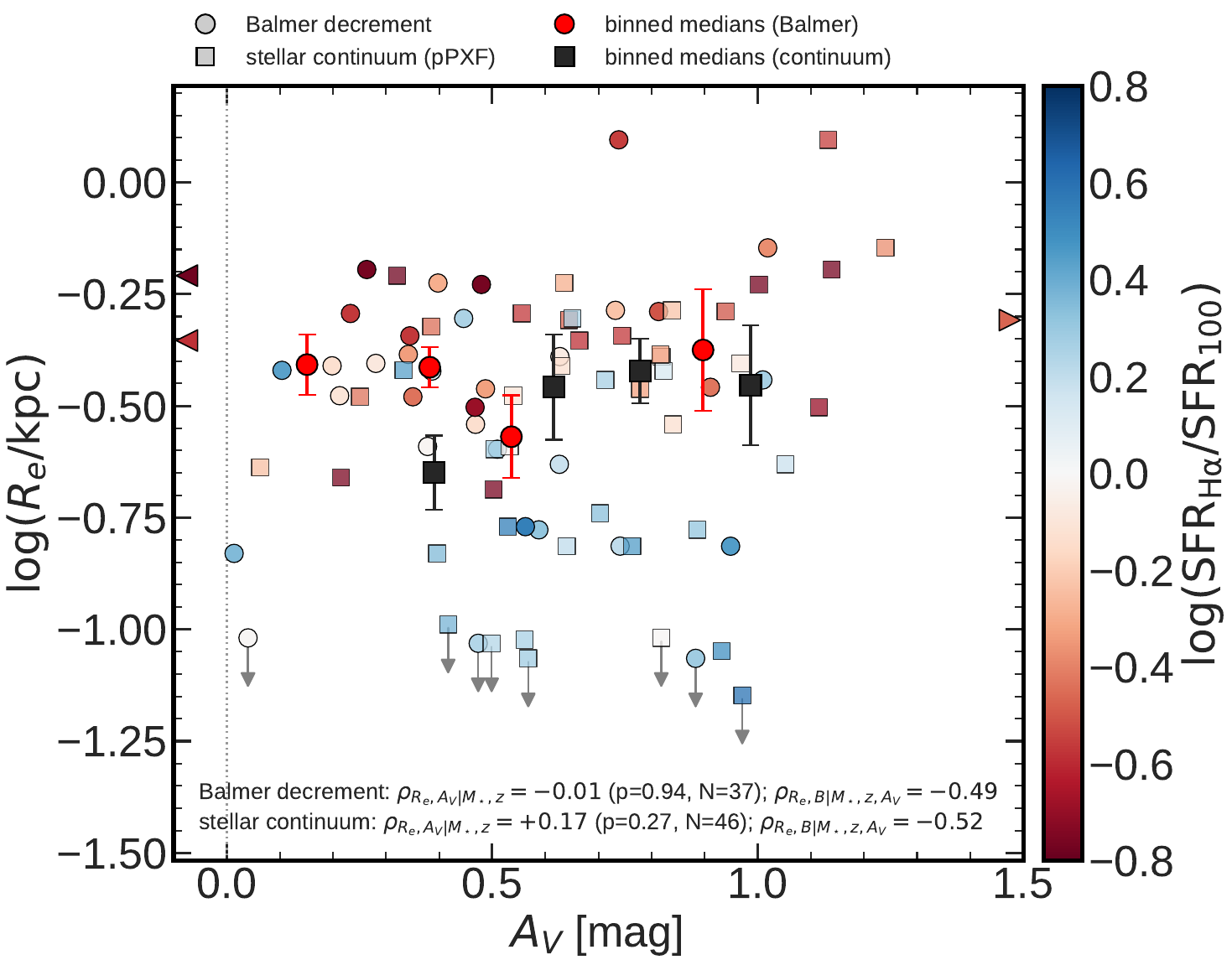}
    \caption{Half-light radius versus dust extinction in the V band for the OASIS sample.
    No strong trend is seen, disfavouring a scenario whereby the observed trends between size and burstiness are caused by dust. The \pccs remain significant between burstiness and size even whilst controlling for stellar mass, redshift and dust.}
    \label{fig:dust-size}
\end{figure}
The importance of the effects of dust in determining galaxy sizes should not be underestimated. One of the effects of dust on galaxy spectra and photometry is to introduce a significant gradient into measurements by reddening the spectra \citep[e.g.][]{Draine2003, Salim2018}. Our interest here is trying to understand if any of our observed trends are likely driven by dust. As an example, many simulations find an inverse size-mass relation at higher redshifts which they believe is driven by a central concentration of dust, unaccounted for in observations \citep{Roper2022, Ceverino2026}. This dust preferentially attenuates the bright UV centres of the galaxies, thereby increasing their observed sizes in the UV. This explanation is required in these simulations in order to match observations that find a clear positive size-mass relation. Other simulations, such as in \citet{McClymont2025}, find that they are able to produce the observed size-mass relations with a more physical multiphase ISM framework including more radiative transfer effects. 

At cosmic noon, dust has been shown to not be responsible for the observed trends of central starbursts and inside out quenching \citep[albeit for much more massive galaxies, e.g.][]{Tacchella2018}, suggesting that at higher redshifts this is also unlikely to be the primary driver of our observed trends. 
Further supporting this argument, comparisons of NUV to FUV sizes at $z=4-6$ find consistency, suggesting a lack of dust gradients in galaxies at $z=4-6$ \citep{Danhaive2026}.

In addition, studies at high-z have found increasingly flatter dust curves \citep[][]{Markov2025, Shivaei2025} with little dust attenuation suggesting that different dust compositions \citep{Langeroodi2024b} or dust removal processes \citep{Lesniewska2025} play an increasingly important role \citep[for a review see][]{Schneider2024}.
All of this conspires to likely lessen the impact of dust on our observed trends, however, it is still important to check for obvious effects. 

In Fig. \ref{fig:dust-size} we explore the effective half-light radius of the \oasis\, sample vs the extinction in the V band ($A_V$) as measured by the Balmer decrement, and also from the stellar continuum colour coded by \sfrratio.
There are naturally a number of caveats in this approach, including galaxies with neither H$\alpha$ or H$\beta$ (not included in the figure), assumption of case B recombination \citep[not always valid][]{McClymont2024}, and the assumption of a chosen attenuation curve. The full details of the dust correction are included in OASIS Paper~I.

Despite these caveats, Fig. \ref{fig:dust-size} shows that we see no strong trend between $A_V$ and size, which means that we are consistent with a mostly flat relation. The \pccs show no significant trend between size and $A_V$, and the relation between size and burstiness remains significant whilst controlling for $A_V$.
This suggests that our observed trends should not be primarily driven by the effects of dust (although we cannot fully rule this out without deep ALMA coverage). A central concentration of dust would serve to redden the centre and increase the measured sizes of the bursting galaxies which is the opposite of what we see here. An alternative explanation whereby dust is expelled from the central region, perhaps via radiation driven outflows \citep{Ferrara2023, Ferrara2024} is still possible, but not necessarily in tension with our central starburst scenario.

\section{ForcePho sizes}
\label{a.s.forcepho}
\begin{figure*}
    \centering
    \includegraphics[width=\linewidth]{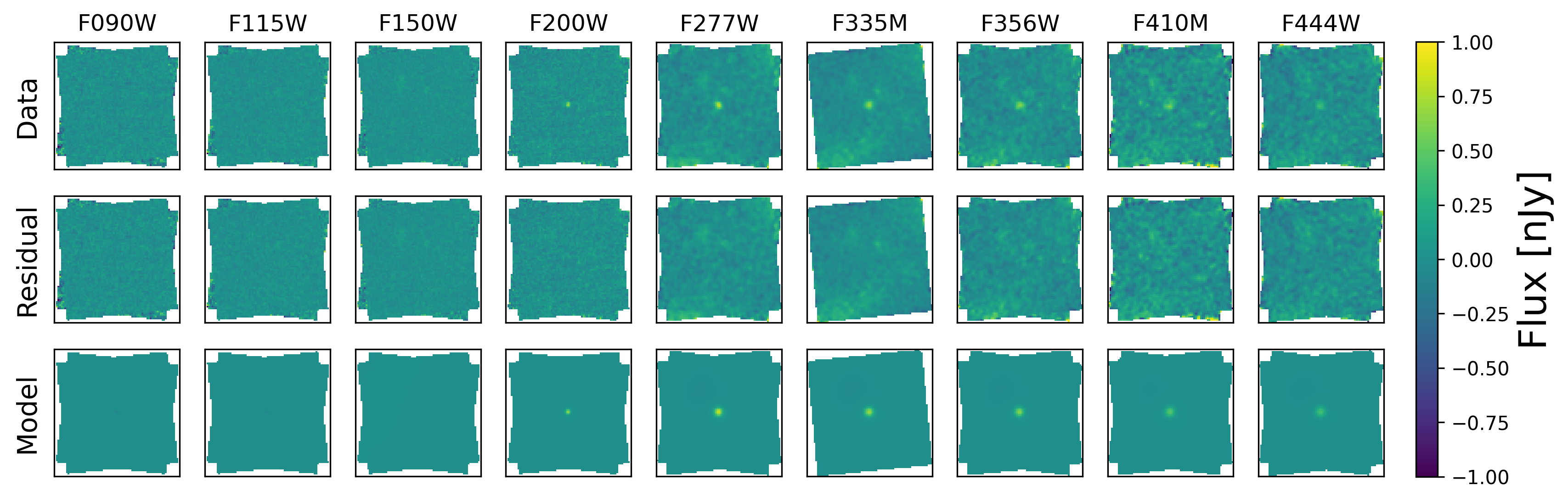}
    \includegraphics[width=\linewidth]{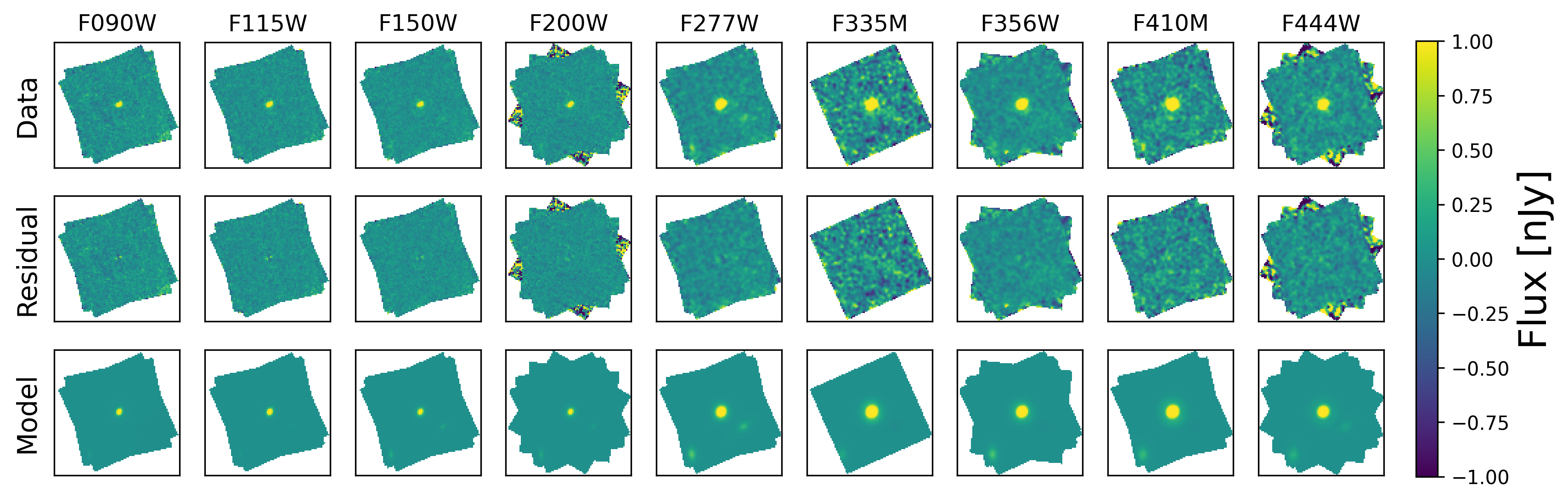}
    \includegraphics[width=\linewidth]{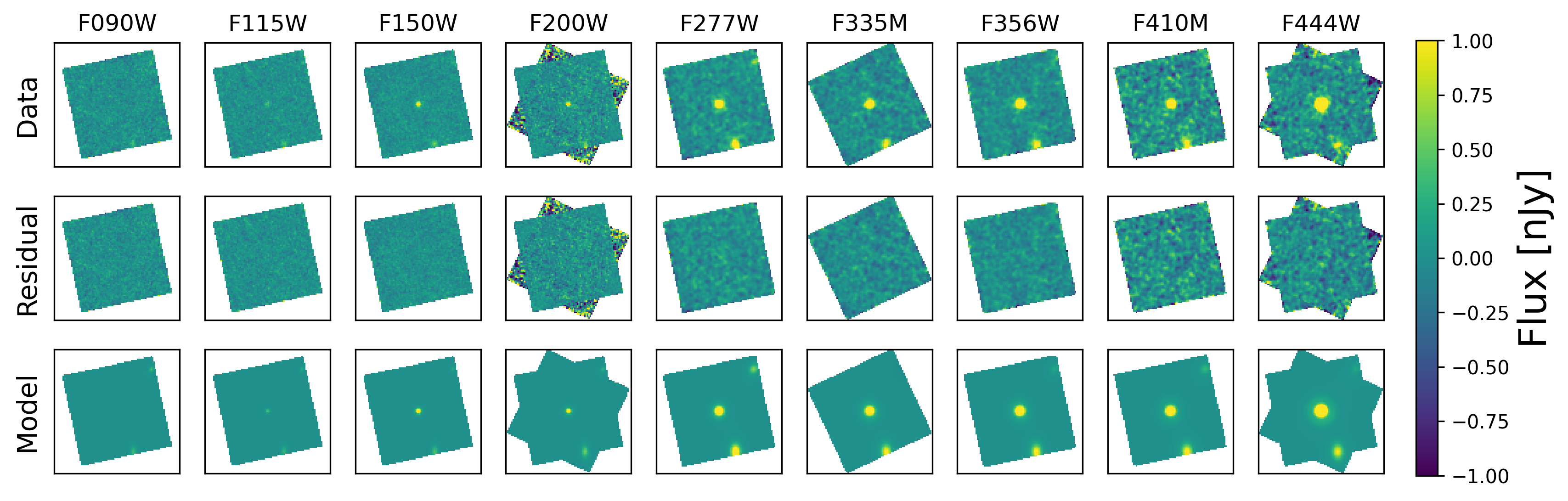}
    \includegraphics[width=\linewidth]{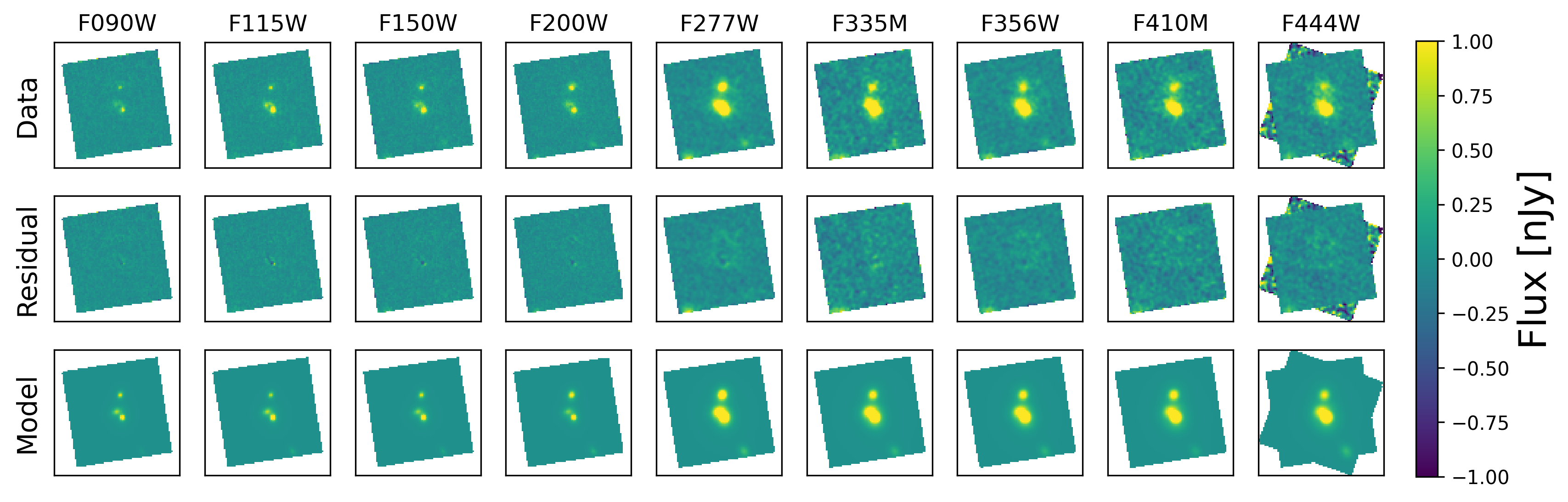}
    \caption{Data, residual model figures for four \forcepho fits. These show that \forcepho well reproduces the observed light distribution within each band even in the cases of multiple sources.}
    \label{fig:pho_fit}
\end{figure*}

\begin{figure*}
    \centering
    \includegraphics[width=\linewidth]{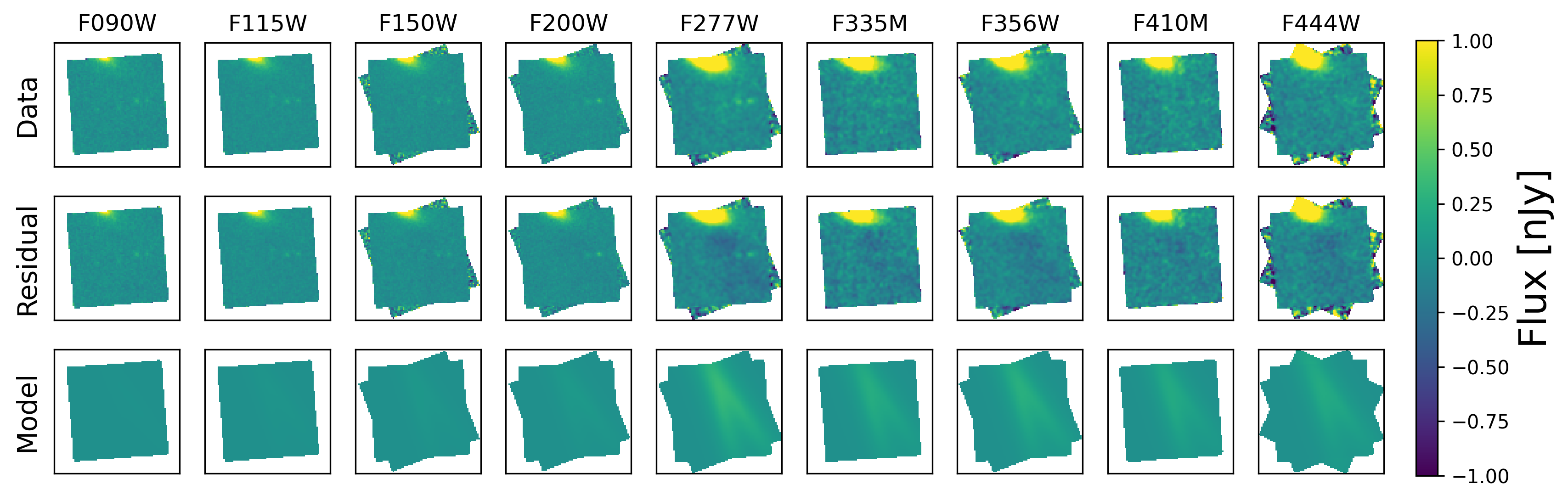}
    \includegraphics[width=\linewidth]{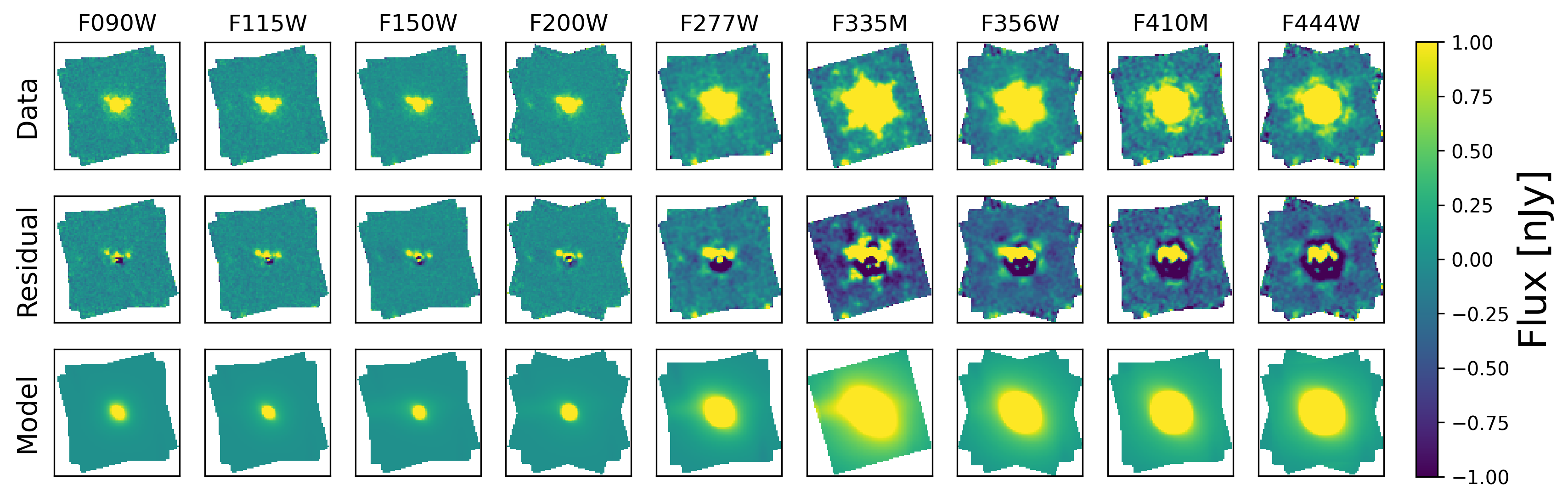}
    \includegraphics[width=\linewidth]{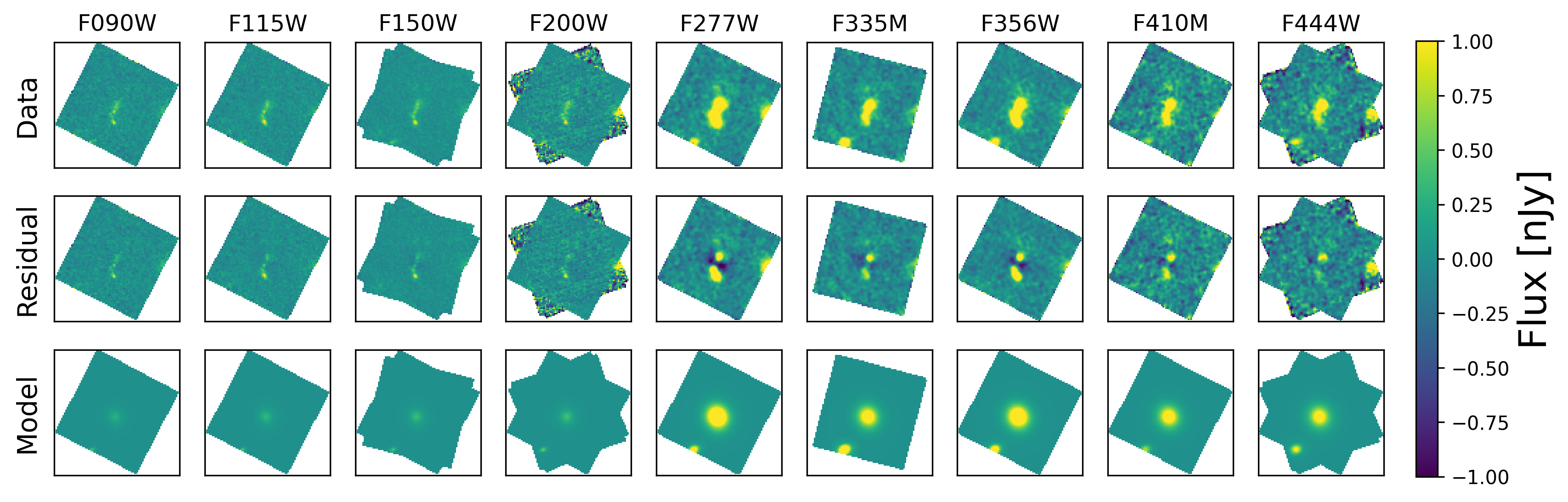}
    \includegraphics[width=\linewidth]{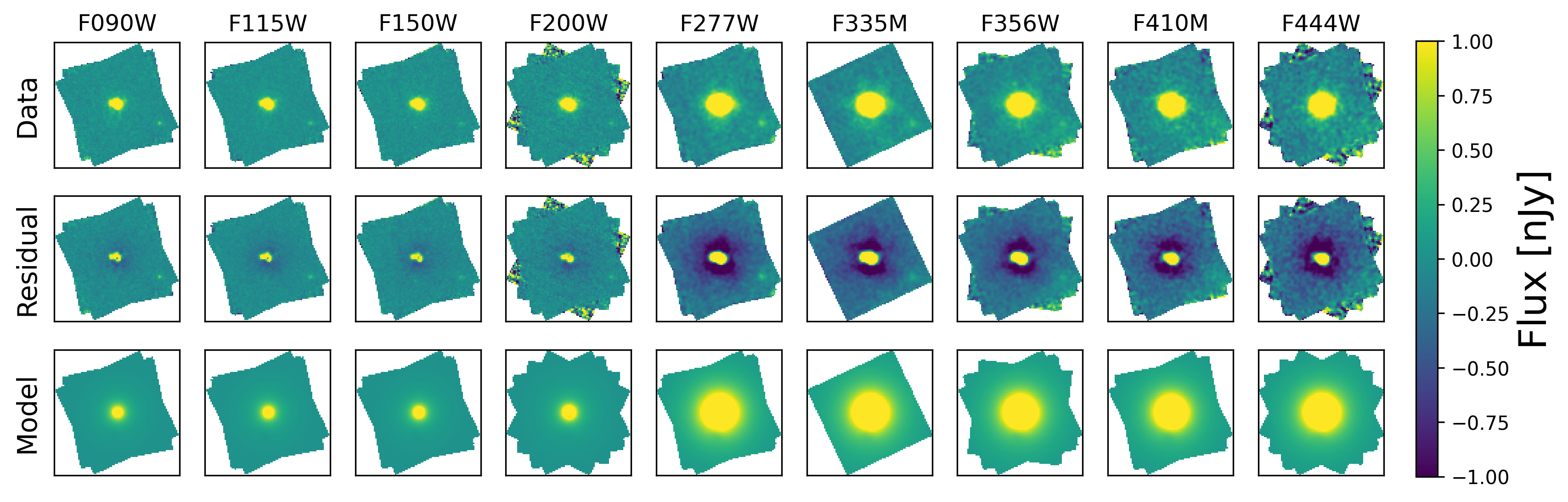}
    \caption{Four of the 5 galaxies that have been removed via visual inspection from the sample. Upper panel and lower panel: galaxy is too faint to be detected in imaging and resides close to bright neighbours. Upper Middle panels: galaxies are extremely large and bright with a complex multi-component structure. Lower: galaxy is extremely large and bright with very high S/N.}
    \label{fig:pho-removed}
\end{figure*}

\begin{figure*}
    \centering
    \includegraphics[width=\linewidth]{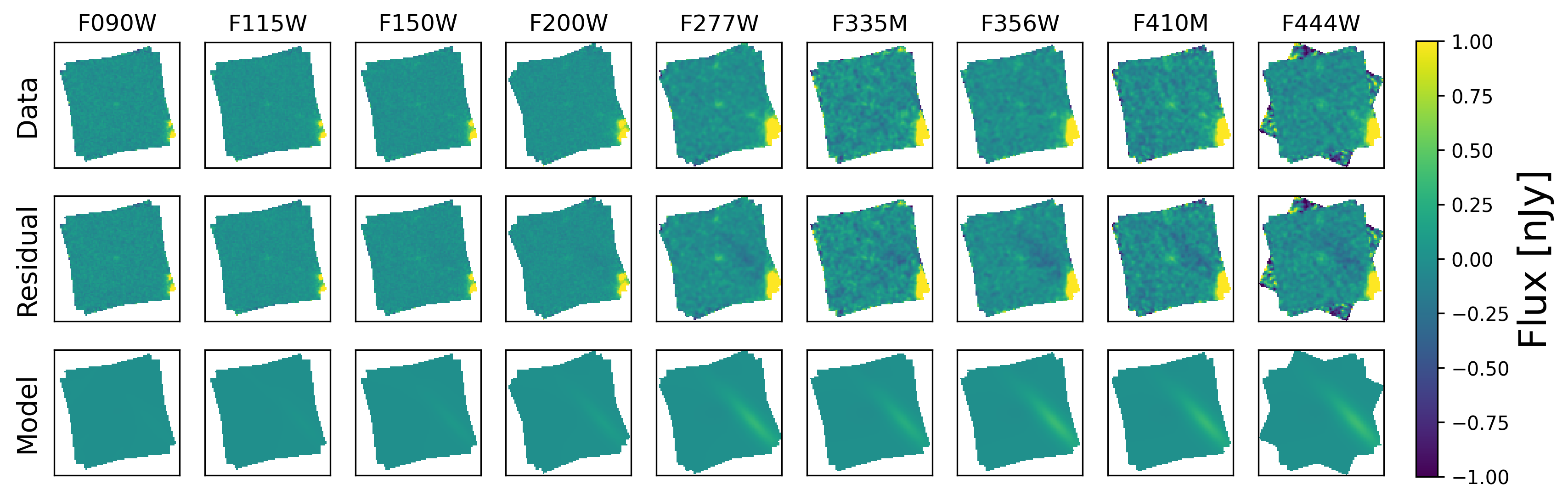}
    \caption{One of the 5 galaxies that have been removed via visual inspection from the sample. }
    \label{fig:pho-removed2}
\end{figure*}

 Fig \ref{fig:pho_fit} shows the data, model and residual stacks from the \forcepho fitting of four representative galaxies from our overall sample. The upper panel shows a faint isolated galaxy, the top middle shows an isolated brighter galaxy, and the middle and lower bottom panels show brighter galaxies with faint and/or bright near neighbours. We fit all galaxies within the individual cutouts. \forcepho fits the individual exposures, enabling the flux distribution to vary between filters but fixing the structural parameters \citep[for more details see][]{Baker2025}. The key aspect is that \forcepho is able to well-reproduce the observed light distribution, both in the case of faint isolated galaxies \citep[e.g.][]{Robertson2023, Tacchella2023} but also in the case of near neighbours \citep[e.g.][]{Baker2025, Carniani2024}.

 However, as previously mentioned in the sample selection section, there were a handful of galaxies that had to be removed via visual inspection due to poor fits. Fig. \ref{fig:pho-removed} shows four examples of these galaxies. The upper panel shows a galaxy that was removed due to no clear detection in the imaging, hence the fit has tried to fit the background resulting in long striped fits. This is the same for the galaxy shown in Fig. \ref{fig:pho-removed2}.
 These are not representative of a galaxy size so these galaxies were removed from the sample. The top middle and bottom panels of Fig. \ref{fig:pho-removed} show two incredibly high S/N large galaxies, whose light distribution has been poorly reproduced by \forcepho. As discussed extensively in previous works \cite{Robertson2023, Tacchella2023, Baker2025} \forcepho is designed for low S/N high-redshift compact objects, hence has significantly struggled with these two very bright targets. However, these targets are not representative of the galaxy population and lie as extreme outliers so were removed from the sample. The lower middle panel shows a complicated multi-component structure, likely undergoing a merger. We removed this due to the poor fit.

\end{appendix}

\end{document}